\documentclass[twocolumn,aps,showpacs,prb,postscript,superscriptaddress]{revtex4-1}
\usepackage{amsmath,amssymb}
\DeclareMathOperator{\Tr}{Tr}
\usepackage{makecell}
\usepackage{subfig}
\usepackage{graphicx}
\graphicspath{{plots/}}
\usepackage{braket}
\usepackage{epstopdf}
\usepackage[usenames]{color}
\usepackage{hyperref}
\usepackage{float}
\usepackage{xcolor}
\usepackage{tikz}
\usetikzlibrary{positioning}
\begin{document}

\title{Transport in the non-ergodic extended phase of interacting quasi-periodic systems}

\author{Soumi Ghosh}
\affiliation{Department of Physics, Indian Institute of Science, Bangalore 560 012, India}
\author{Jyotsna Gidugu}
\affiliation{Department of Physics, Indian Institute of Science, Bangalore 560 012, India}
\author{Subroto Mukerjee }
\affiliation{Department of Physics, Indian Institute of Science, Bangalore 560 012, India}

\begin{abstract}
We study the transport properties and the spectral statistics of a one-dimensional closed quantum system of interacting spinless fermions in a quasi-periodic potential which produces a single particle mobility edge in the absence of interaction. For such systems, it has been shown that the many body eigenstates can be of three different kinds: extended and ETH (eigenstate thermalization hypothesis) obeying (thermal), localized and ETH violating (many body localized) and extended and ETH violating (non-ergodic extended). Here we investigate the non-ergodic extended phase from the point of view of level spacing statistics and charge transport. We calculate the dc conductivity and the low frequency conductivity $\sigma(\omega)$ and show that both are consistent with sub-diffusive transport. This is contrasted with diffusive transport in the thermal phase and blocked transport in the MBL phase.
\end{abstract}

\maketitle
\section{Introduction:}
Anderson in his seminal paper (1958) \citep{Anderson.1958} showed that sufficiently strong disorder can localize all the single particle states in a disordered system. An extension of this phenomenon in the presence of interactions called many body localization (MBL) was later argued to occur \citep{baa.2006,OganesyanHuse.2007}. Since then, it has been a subject of immense interest and there have been various numerical\citep{mbl_numerical_znidaric, PalHuse.2010, mbl_numerical_bardarson}, phenomenological \citep{phenomenological_serbyn, phenomenological_huse} and analytical \citep{analytical_imbrie} studies along with experimental verification \citep{Bloch.2015, SPME_Bloch.2019}.
Many body localized systems are isolated interacting many body systems which generically fail to thermalize\citep{OganesyanHuse.2007,PalHuse.2010}, violating the eigenstate thermalization hypothesis (ETH)\citep{deutsch.1991,srednicki.1994,Rigol.2008}. As a consequence of the violation of ETH, the bipartite entanglement entropy of these systems in energy eigenstates is not equal to the thermal entropy. In fact the entanglement in energy eigenstates follows an area law behavior in the MBL phase~\citep{QPMBL.2013}. It has been argued that these systems have emergent conservation laws~\citep{Huse.Pheno.2014,modak1.2015,Serbyn.2013,Chandran.2015} which prevents thermalization in a way similar to the non-ergodic behavior seen in the integrable systems~\citep{Integrable.Rigol, Chaos.Rigol,Integrable.Modak,Modak.2014}. As a consequence of localization there is no diffusion in the system. In the presence of interactions, for small enough disorder, generic many body systems remain in a thermal phase. However, as the disorder strength is increased, for sufficiently large disorder, the system can undergo a thermal-MBL phase transition, which is a dynamical phase transition unlike the equilibrium phase transitions widely studied in physics. This thermal-MBL transition has been also shown to exist in the presence of a quasi-periodic potential instead of true random disorder \citep{QPMBL.2013,Bloch.2015}.

The simplest quasi-periodic potential is the Aubry-Andre potential ($h_i=h \cos(2\pi \alpha i+\phi)$) in the presence of which the non-interacting system has extended or localized single particle states depending on the potential strength, $h$. \citep{AA.model} For $h<2t$ all the single particle energy eigenstates are extended and for $h>2t$ all the energy eigenstates are localized where $t$ is the hopping parameter. Upon switching on interactions, this delocalized to localized transition transforms into a thermal-MBL transition where for $h<h_c$ all typical mid spectrum states are thermal and for $h>h_c$ the typical states are many body localized~\citep{QPMBL.2013}.

However there are other quasiperiodic potentials\citep{quasiperiodic_spme_1, quasiperiodic_spme_2, quasiperiodic_spme_3} which have single particle mobility edges (SPME) in the absence of interactions upto some critical potential strength, and the SPME is defined by a critical energy $E_c$ that separates the extended and localized eigenstates of the single particle Hamiltonian. In fact, single particle mobility edges arise generically for quasiperiodic potentials. The Aubry-Andre model is in a sense fine-tuned due to which it possesses an energy independent duality which does not allow the existence of a mobility edge.

In the presence of interactions these models also undergo transitions from thermal to athermal phases~\citep{modak_mukerjee_SPME, modak_ghosh_mukerjee}. However, it has been argued that these models are different from those without single particle mobility edges in that they possess a non-ergodic extended phase at the intermediate potential strengths. In this phase, there are states in the energy spectrum that are non-ergodic in the sense that they violate the ETH but are also extended since they possess volume law satisfying entanglement entropy~\citep{LiGaneshan.2015,Modak.Mukerjee.Review}. It has been argued \citep{LiGaneshan.2015} that this phase can be thought of as arising from the many-body states of the non-interacting model which are Slater determinants with both extended and localized states upon turning on interactions ``adiabatically''. This kind of non-ergodic state was also described in Josephson junction arrays (JJA)~\citep{Pino.2016} where it was shown that there is a phase transition as a function of temperature where the metallic phase at low temperature and the many body localizing phase at higher temperatures were separated by a `bad metal' phase where the system is non-ergodic but conducting.

In this paper we investigate the transport properties of the non-ergodic extended phase in systems with single particle mobility edge. Using numerical exact diagonalization, we obtain the energy level spacing statistics and the dc and ac conductivities, in terms of which we characterize the phase. Here we consider spinless fermions in one dimension described by the Hamiltonian
\begin{equation}\nonumber
H=-t \sum\limits_{i}^{} \left(c_{i+1}^{\dagger} c_{i}^{} + h.c. \right) + \sum\limits_i h_i n_i + V\sum\limits_i n_i n_{i+1}
\end{equation}
where $t$ is the hopping parameter, $V$ is the nearest neighbor interaction strength and the onsite potential $h_i$ is chosen to be the quasiperiodic potential $h_i=h\frac{\cos(2\pi\alpha i+\phi)}{1-\beta \cos(2\pi\alpha i+\phi)}$ which has a mobility edge \citep{quasiperiodic_spme_3} at $\beta E_c=2 {\rm sgn} (h)(|t|-|h|/2)$ separating the single particle extended states from the localized ones.
We calculate the level statistics and the transport properties numerically using exact diagonalization for different filling fractions ($\nu=1/2,1/4,1/6$). For better statistics we average all the quantities over randomly chosen offset angle $\phi \in\left[ 0, 2\pi \right)$ and apply periodic boundary conditions in all our calculations. We calculate the variation of average level spacing ratio as a function of energy density  to locate a critical energy density separating the energy eigenstates with Poisson energy level spacing statistics from the states with the energy level spacing following a Gaussian Orthogonal Ensemble (GOE) distribution. We also examine the violation of ETH and obtain the entanglement scaling as a function of energy density at these different filling fractions to locate transitions from a non-ergodic to ergodic phase and localized to delocalized phase respectively in the energy spectrum~\citep{LiGaneshan.2015}. We obtain the critical values of energy density at which these transitions occur.
\begin{figure*}[t]
	\centering
	\begin{tikzpicture}
	\node (img1){\includegraphics[width=0.35\textwidth]{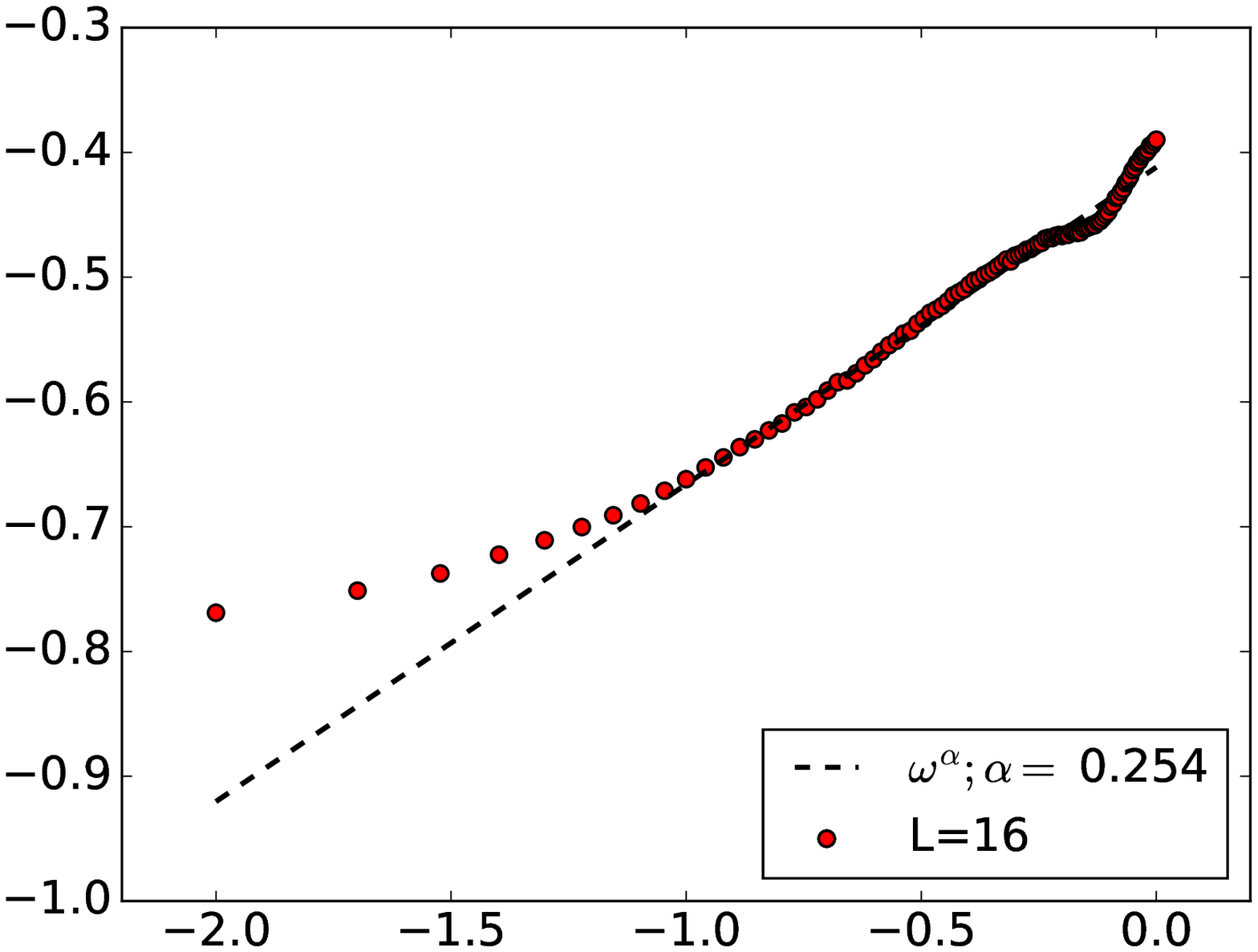}};
	\node (img2) [right= of img1,node distance=0cm,xshift=-1.5cm]{\includegraphics[width=0.35\textwidth]{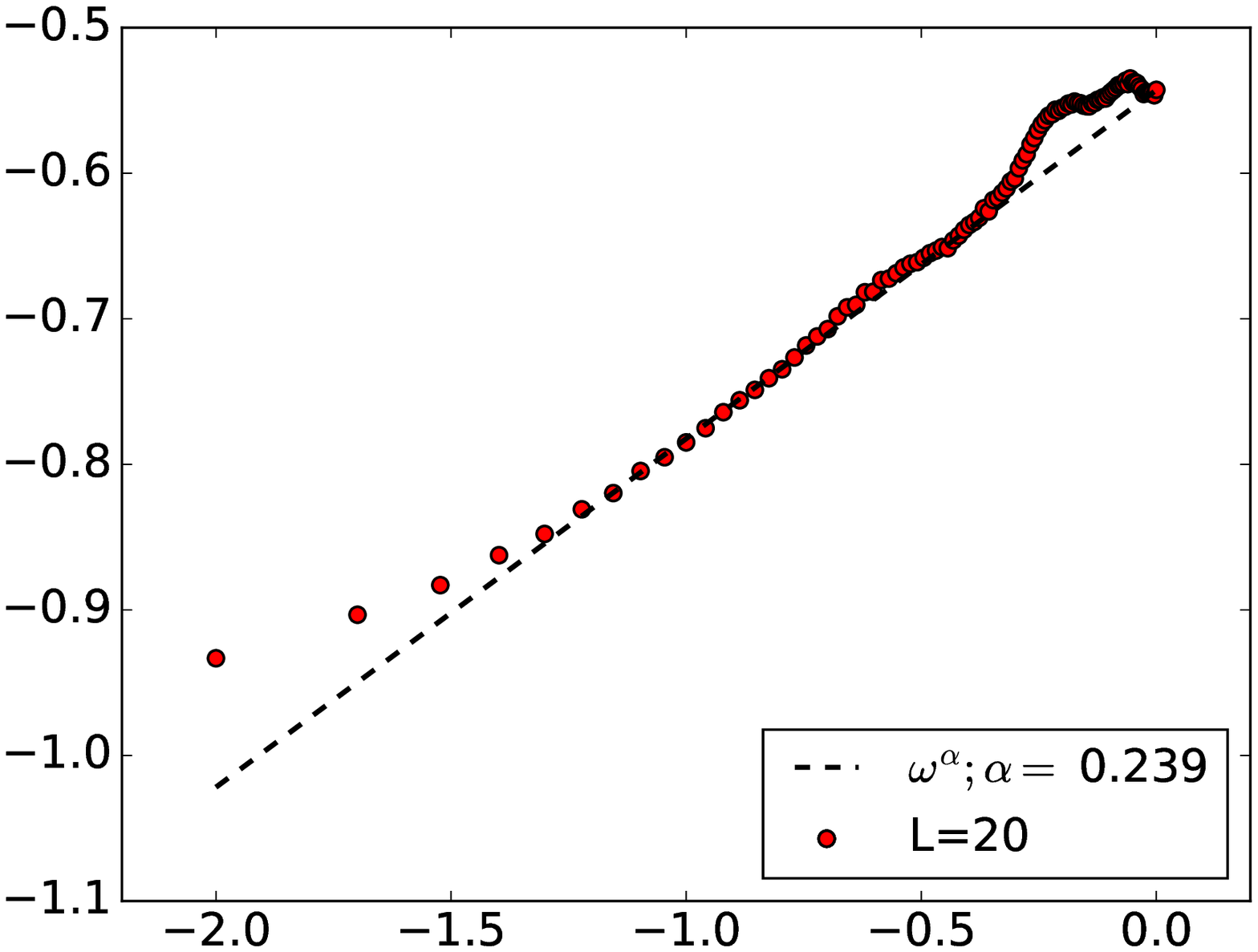}};
	\node(img3) [right= of img2, node distance=0cm,xshift=-1.5cm] {\includegraphics[width=0.35\textwidth]{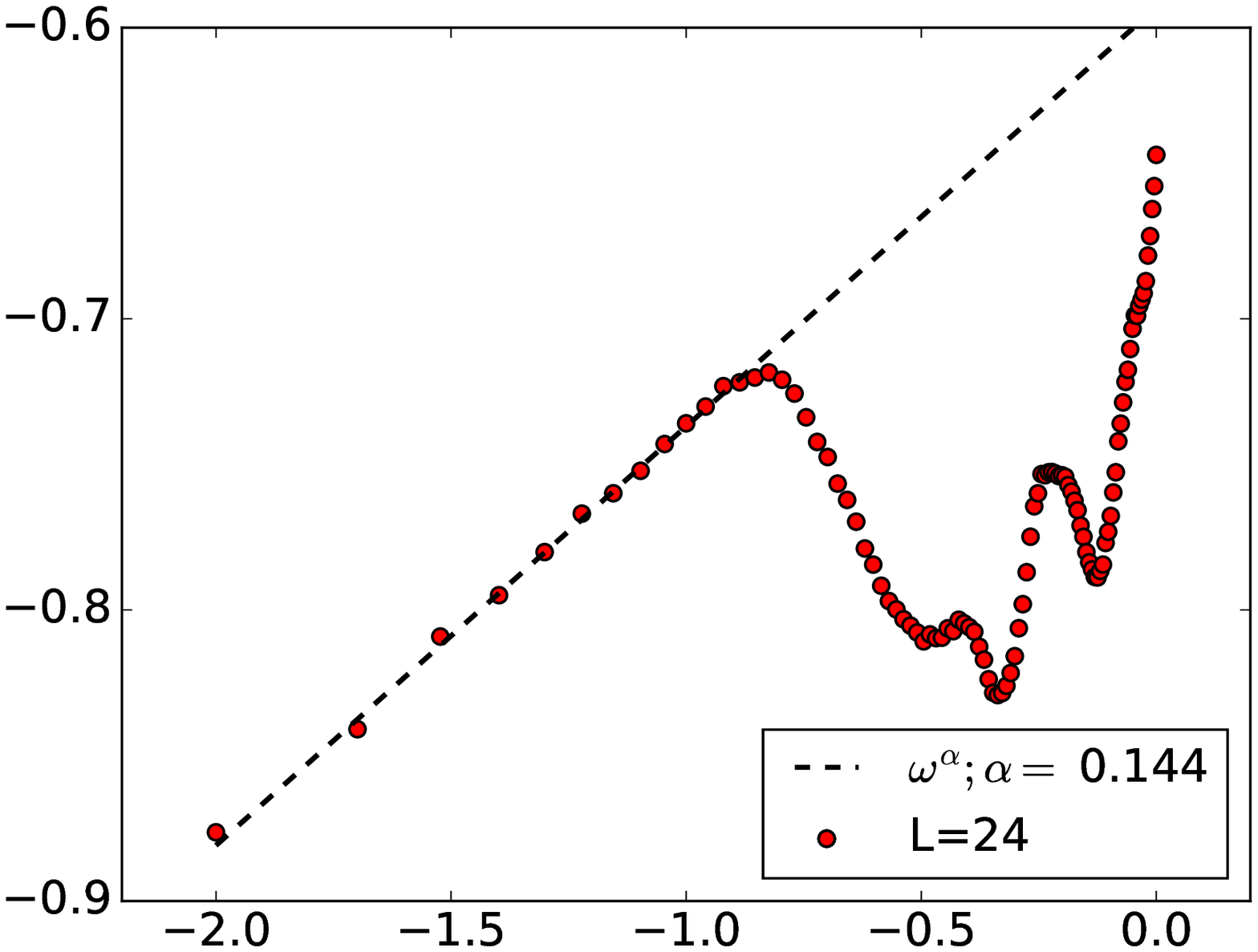}};
	\node[left=of img1,node distance=0cm,yshift=1.0cm,xshift=1.0cm,rotate=90]{$log\left[\sigma(\omega)\right]$};
	\node[below=of img1,node distance=0cm,yshift=1.4cm]{$log(\omega)$};
	\node[below=of img2,node distance=0cm,yshift=1.4cm]{$log(\omega)$};
	\node[below=of img3,node distance=0cm,yshift=1.4cm]{$log(\omega)$};

	\node[above= of img1,node distance=0cm,yshift=-3.0cm,xshift=-1.0cm]{\fbox{$\nu=1/2$}};
	\node[above= of img2,node distance=0cm,yshift=-3.0cm,xshift=-1.0cm]{\fbox{$\nu=1/4$}};
	\node[above= of img3,node distance=0cm,yshift=-3.0cm,xshift=-1.0cm]{\fbox{$\nu=1/6$}};

	\node[above= of img1,node distance=0cm,yshift=-2.1cm,xshift=-2.0cm]{$(a)$};
	\node[above= of img2,node distance=0cm,yshift=-2.1cm,xshift=-2.0cm]{$(b)$};
	\node[above= of img3,node distance=0cm,yshift=-2.1cm,xshift=-2.0cm]{$(c)$};
	\end{tikzpicture}
	\caption{Log-log plot of $\sigma(\omega)$ vs $\omega$: This plot shows the variation of $\sigma(\omega)$ as a function of $\omega$ for different filling fractions $\nu=\frac{1}{2}~(a),\frac{1}{4}~(b),\frac{1}{6}~(c)$ at system sizes $L=16,20$ and $24$ respectively. The dashed lines are fit for the power law $\sigma(\omega)\sim \omega^\alpha$ at frequencies well below the hopping parameter $t$. The fitted parameter $\alpha$ belong to the sub-diffusive range for all filling fractions. The width $\eta$ of the Lorentzian (broadening of the delta function in Kubo formula) is chosen to be 0.1$\Delta$ where $\Delta$ is the mean level spacing.}
	\label{ac}
\end{figure*}

We also calculate the low frequency optical (AC) conductivity. It has been shown~\citep{kagarwal.2015} that in the disordered $XXZ$ model, the AC conductivity vanishes at low frequencies as a power law $\sigma(\omega) \sim \omega^\alpha$ where the parameter $\alpha \rightarrow 1$ as the localized phase is approached and $\alpha \rightarrow 0$ in the diffusive phase. It was also shown that there exists a sub-diffusive phase in the thermal region near the thermal-MBL transition where $0<\alpha<1$. This sub-diffusive phase is associated with Griffiths effects where rare local regions of strong disorder in a random one dimensional system act as insulating regions and restrict transport in the system.
In our quasi-periodic system we investigate numerically the infinite temperature low frequency behavior of the ac conductivity $\sigma(\omega)$ and the system size dependency of the dc conductivity. In this paper we show that in the presence of a non-ergodic extended phase, the AC conductivity vanishes at low frequency following a power law $\sim \omega^\alpha$ where $\alpha$ has a value between $0$ and $1$ for different filling fractions indicating subdiffusive behavior and that the dc conductivity decreases with the increasing system sizes but decay slower than an exponential decay expected in MBL phase~\citep{dc_cond_setiawan}, indicating subdiffusive transport. We also calculate the individual contributions to the dc conductivity from the three different phases in the energy spectrum, namely the many-body localized, non-ergodic extended and ergodic phase to extract the information about the scaling of the same with increasing system sizes and find that the contribution from the non-ergodic extended phase decays slowly than the contribution from the many body localized phase while the contribution from the ergodic phase hardly decays at all with the increasing system sizes.
\section{Conductivity calculations:}
To study the charge transport within the system subject to periodic boundary conditions we use the Kubo formula for the conductivity $\sigma (\omega)$, 
\begin{equation}\nonumber
\sigma(\omega)=\dfrac{\pi}{L}\dfrac{1-e^{\beta \omega}}{\omega}\sum\limits_{m,n} e^{-\beta E_m} |\braket{m|J|n}|^2 \delta\left(E_n-E_m-\omega\right)
\end{equation}
where $J=-it\sum\limits_i (c_{i+1}^\dagger c_i^{}-c_i^\dagger c_{i+1}^{})$ is the total current operator and $\ket{n}$ and $\ket{m}$ are many body eigenstates of the system.

In the study of many body localization, localization properties of the closed quantum systems are typically studied at infinite temperature. This is because the many body density of states is sharply peaked in the middle of the spectrum for a sufficiently large system. Hence if all energy eigenstates are chosen with equal weight, which corresponds to putting the system in contact with an infinite temperature ($\beta \rightarrow 0$) bath, any quantity calculated by averaging over the whole energy spectrum gives the typical mid spectrum value of that quantity. The most convenient way to calculate the conductivity is to use the Kubo formula, which has temperature as a parameter. Thus, to obtain the contributions to the conductivity from all the energy eigenstates in an unbiased manner, the temperature is set to infinity for our calculations. At infinite temperature, the Kubo formula reduces to
\begin{equation}
T\sigma(\omega)=\dfrac{\pi}{ZL}\sum\limits_{m,n} |\braket{m|J|n}|^2 \delta\left(E_n-E_m-\omega\right)
\label{eq-ac}
\end{equation}
where $Z$ is the partition function.
The many body energy spectrum is discrete for finite sized systems and approaches a continuous spectrum in the thermodynamic limit. The discreteness of the energy spectrum results in the conductivity being a sum of discrete delta functions rather than the smooth function of $\omega$ expected in the thermodynamic limit. Thus, for a finite-sized system, the delta functions are approximated by Lorentzian function of the form $\delta(E)=\dfrac{1}{\pi}\dfrac{\eta}{\eta^2+E^2}$, where the width of the Lorentzian peak $2\eta$ is chosen such that $\eta<< \Delta$. This allows one to obtain a smooth form for $\sigma (\omega)$. Here $\Delta$ is the mean level spacing of the many body spectrum.
\begin{figure*}[t]
	\centering
	\begin{tikzpicture}
	\node (img1){\includegraphics[width=0.35\textwidth]{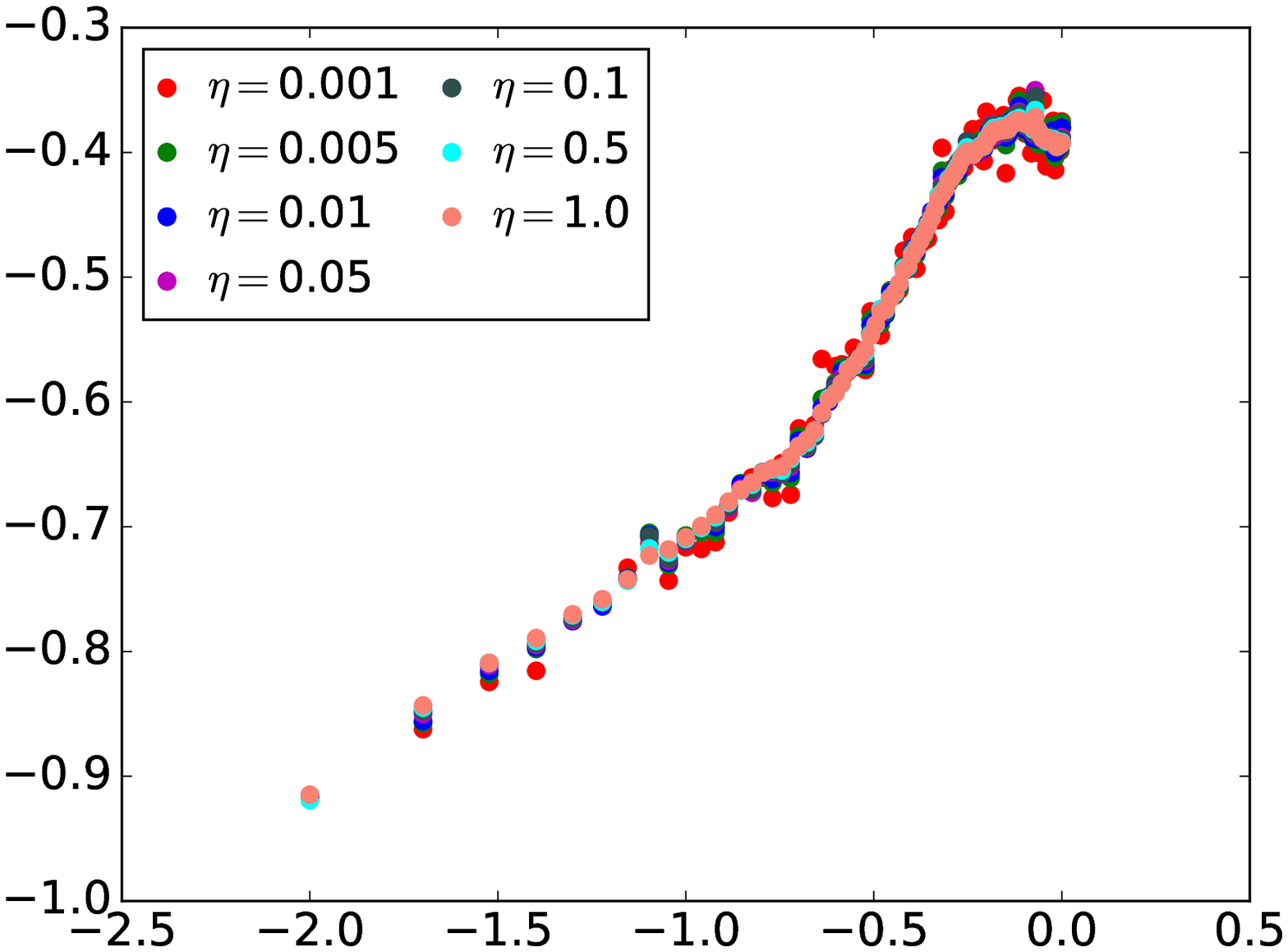}};
	\node (img2) [right= of img1,node distance=0cm,xshift=-1.5cm] {\includegraphics[width=0.35\textwidth]{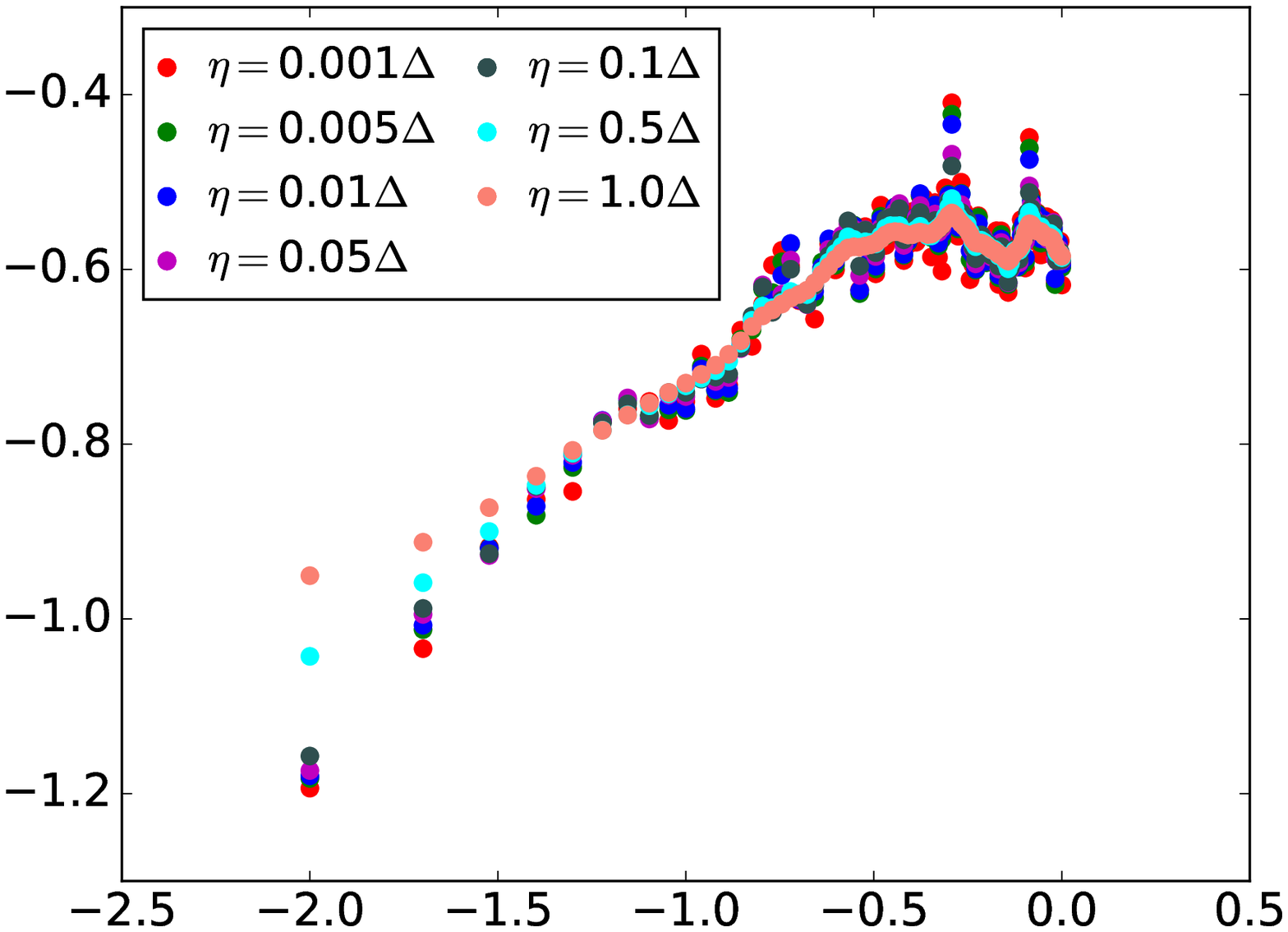}};
	\node(img3) [right= of img2, node distance=0cm,xshift=-1.5cm] {\includegraphics[width=0.35\textwidth]{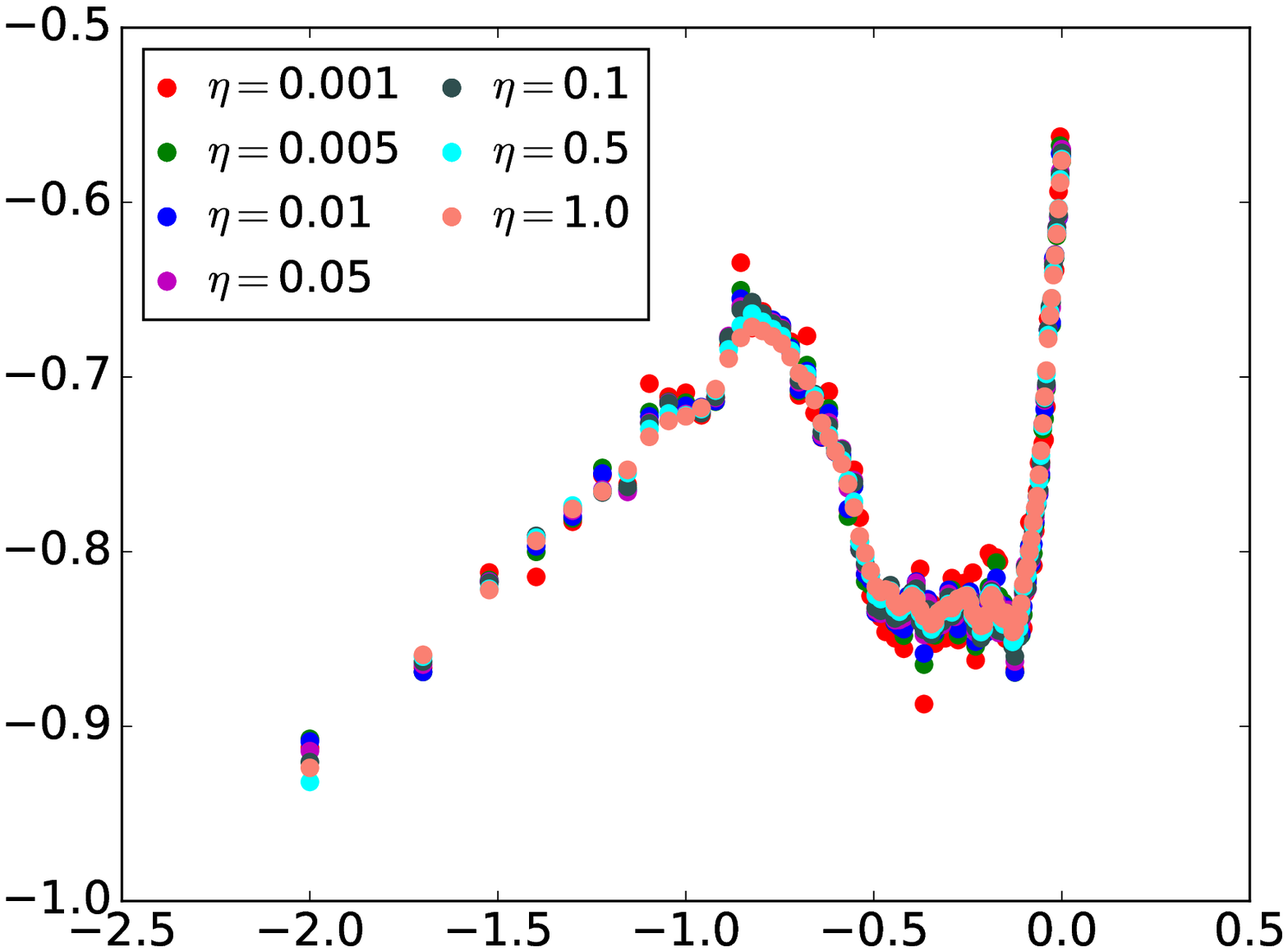}};
	\node[left=of img1,node distance=0cm,yshift=1.0cm,xshift=1.0cm,rotate=90]{$log\left[\sigma(\omega)\right]$};
	\node[below=of img1,node distance=0cm,yshift=1.4cm]{$log(\omega)$};
	\node[below=of img2,node distance=0cm,yshift=1.4cm]{$log(\omega)$};
	\node[below=of img3,node distance=0cm,yshift=1.4cm]{$log(\omega)$};

	\node[above= of img1,node distance=0cm,yshift=-3.4cm,xshift=-1.2cm]{$\nu=1/2$};
	\node[above= of img2,node distance=0cm,yshift=-3.4cm,xshift=-1.2cm]{$\nu=1/4$};
	\node[above= of img3,node distance=0cm,yshift=-3.4cm,xshift=-1.2cm]{$\nu=1/6$};

	\node[above= of img1,node distance=0cm,yshift=-2.1cm,xshift=2.2cm]{$(a)$};
	\node[above= of img2,node distance=0cm,yshift=-2.1cm,xshift=2.2cm]{$(b)$};
	\node[above= of img3,node distance=0cm,yshift=-2.1cm,xshift=2.2cm]{$(c)$};
	\end{tikzpicture}
	\caption{$\sigma(\omega)$ vs $\omega$ plot for different choices of the Lorentzian width: This plot shows the variation $\sigma(\omega)$ vs. $\omega$ curve with the change of broadening width $\eta$ of the Lorentzian for different filling fractions $\nu=\frac{1}{2}~(a),\frac{1}{4}~(b),\frac{1}{6}~(c)$ at system sizes $L=12,12,18$ respectively. Here $\eta$ is chosen as some fraction $f$ of the mean level spacing $\Delta$ where $f$ is changed from $0.001$ to $1.0$ for all the three cases.}
	\label{ac-eta}
\end{figure*}
\subsection{Low frequency conductivity:}
Here we calculate the low frequency conductivity $\sigma(\omega) \sim \omega^\alpha$ and obtain the exponent $\alpha$ for different values of the filling fraction $\nu=\frac{1}{2},\frac{1}{4},\frac{1}{6}$ for a typical set of parameters, $h=0.6, V=1.0, t=1.0$. The conductivity $\sigma(\omega)$ is calculated by averaging over $1024$ randomly chosen offset $\phi$'s for all the filling fractions. 

It is known that \citep{low_freq_cond_exponent} near the MBL transition $\alpha \rightarrow 1$ and it remains between $1$ and $2$ in the MBL phase. In the thermal phase the conductivity is expected to be diffusive which means that $\sigma(\omega)$ at small $\omega$ should tend to $\sigma_{dc} \neq 0$ and the exponent $\alpha=0$. In \citep{kagarwal.2015}, a sub-diffusive phase with $0<\alpha<1$ was reported signaling the onset of Griffiths effects in a system with random disorder. 

Fig.~\ref{ac} shows the variation of $\sigma(\omega)$ for small values of $\omega<t$. We fit a power law $\sim \omega^\alpha$ for $\omega > \Delta$, $\Delta$ being the mean level spacing. For all the filling fractions $\nu$ we see a value $\alpha$ corresponding to the sub-diffusive transport. We note that there are no Griffiths effects in our systems since the onsite potential is completely deterministic. We thus attribute the sub-diffusive transport to the presence of the non-ergodic extended phase in these systems as discussed further in the following sections.
\paragraph*{Choice of $\eta$:}
Fig.~\ref{ac-eta} shows the variation of $\sigma(\omega)$ versus $\omega$ curve as we change the width of the Lorentzian $2\eta$ for the system sizes $L=12,12,18$ at fillings $\nu=\frac{1}{2},\frac{1}{4},\frac{1}{6}$ respectively.
Here the sample sizes chosen are $4800, 6400, 4800$ respectively for the three filling factions($1/2,1/4,1/6$).
As we see, only the $\nu=\frac{1}{4}$ case has any significant variation as $\eta$ is changed from $0.001\Delta$ to $\Delta$, where $\Delta$ is the mean level spacing. This can be understood by comparing the range of $\omega$ considered for $\sigma(\omega)$ and the value of the mean level spacing $\Delta$ in each cases. For $\nu=\frac{1}{4}$ at $L=12,\:p=3$, the mean level spacing $\Delta \sim 0.05$; while for $\nu=\frac{1}{2}$ and $\frac{1}{6}$ both $\Delta \sim 0.01$ at $L=12$ and $18$ respectively. Now, let us suppose we approximate the delta function $\delta(E_n-E_m-\omega)$ by a box of width $2\eta$ around $\omega=E_n-E_m$ instead of the Lorentzian approximation used in the numerical calculation. Then the energy pairs $(E_n,E_m)$ contributing $|\braket{n|J|m}|^2$ to $\sigma(\omega)$ are given by the condition
\begin{equation}
  \nonumber
  \omega-\eta \leq (E_n-E_m) \leq \omega +\eta
\end{equation}
Therefore when $\eta << \omega$, which is satisfied when $\eta$ is chosen as some fraction of $\Delta$ and $\Delta$ is less than the range of $\omega$ considered ($\eta<\Delta<\omega$), the number of pairs $(E_n,E_m)$ contributing to $\sigma(\omega)$ does not depend on the choice of $\eta$. However, when $\omega \lesssim \Delta$ which is true in Fig.~\ref{ac-eta}[b.], with decreasing $\eta$ (which is again some fraction of $\Delta$), fewer pairs contribute to $\sigma(\omega)$ resulting in the strong dependence of the $\sigma(\omega)$ versus $\omega$ curves on the choice of $\eta$. For the same reason, going to a larger system size and thus decreasing $\Delta$ further, one can get a significant range of $\omega<t$ where the curve is independent of the choice of $\eta$ and  the fitting can be done. Hence we fit for the parameter $\alpha$ in fig.~\ref{ac} at the maximum accessible system sizes for different filling fractions.
\begin{figure}[t]
	\centering
	\begin{tikzpicture}
	\node (img1){\includegraphics[width=0.50\textwidth]{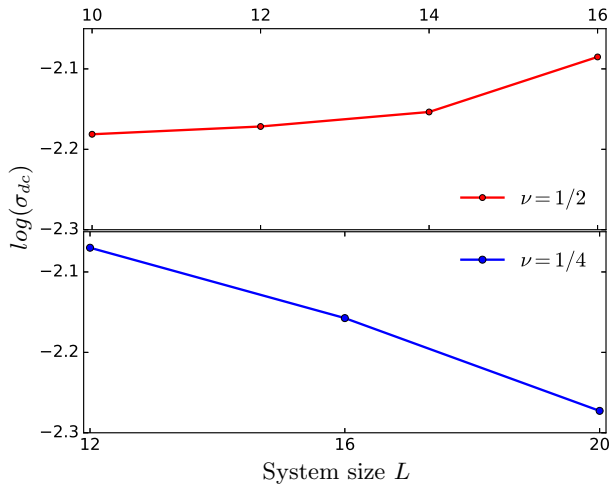}};
	\node[left=of img1,node distance=0cm,yshift=1.0cm,xshift=1.4cm,rotate=90]{$log(\sigma_{dc})$};
	\node[below=of img1,node distance=0cm,yshift=1.5cm]{System size $L$};
	\end{tikzpicture}
	\caption{Variation of the dc conductivity with system size: This shows the variation of $\sigma_{dc}$ with increasing system sizes at filling fractions $\nu=1/2$(Red line in top panel) and $\nu=1/4$(Blue line in bottom panel).The width $\eta$ of the Lorentzian is chosen to be $0.01\Delta$, where $\Delta$ is the mean level spacing.}
	\label{dc}
\end{figure}
\subsection{d.c. conductivity:}
We also calculate the dc conductivity by taking the limit $\omega \rightarrow 0$ in eqn.~\ref{eq-ac}.
\begin{equation}
\lim_{\omega\to 0} T\sigma(\omega)=\dfrac{\pi}{ZL}\sum\limits_{m,n} |\braket{m|J|n}|^2 \delta\left(E_n-E_m\right)
\label{eq-dc}
\end{equation}
A diffusive system must have a non-zero dc conductivity while many-body localized systems are expected to have no transport at all which implies zero dc conductivity. However finite size systems will have some non-zero value of the dc conductivity and for MBL systems this dc conductivity decays exponentially with increasing system sizes implying no transport in the thermodynamic limit. \citep{dc_cond_mbl_1, dc_cond_mbl_2, dc_cond_setiawan}
\begin{figure}[t]
	%\centering
	\begin{tikzpicture}
	\node (img1){\includegraphics[width=0.5\textwidth]{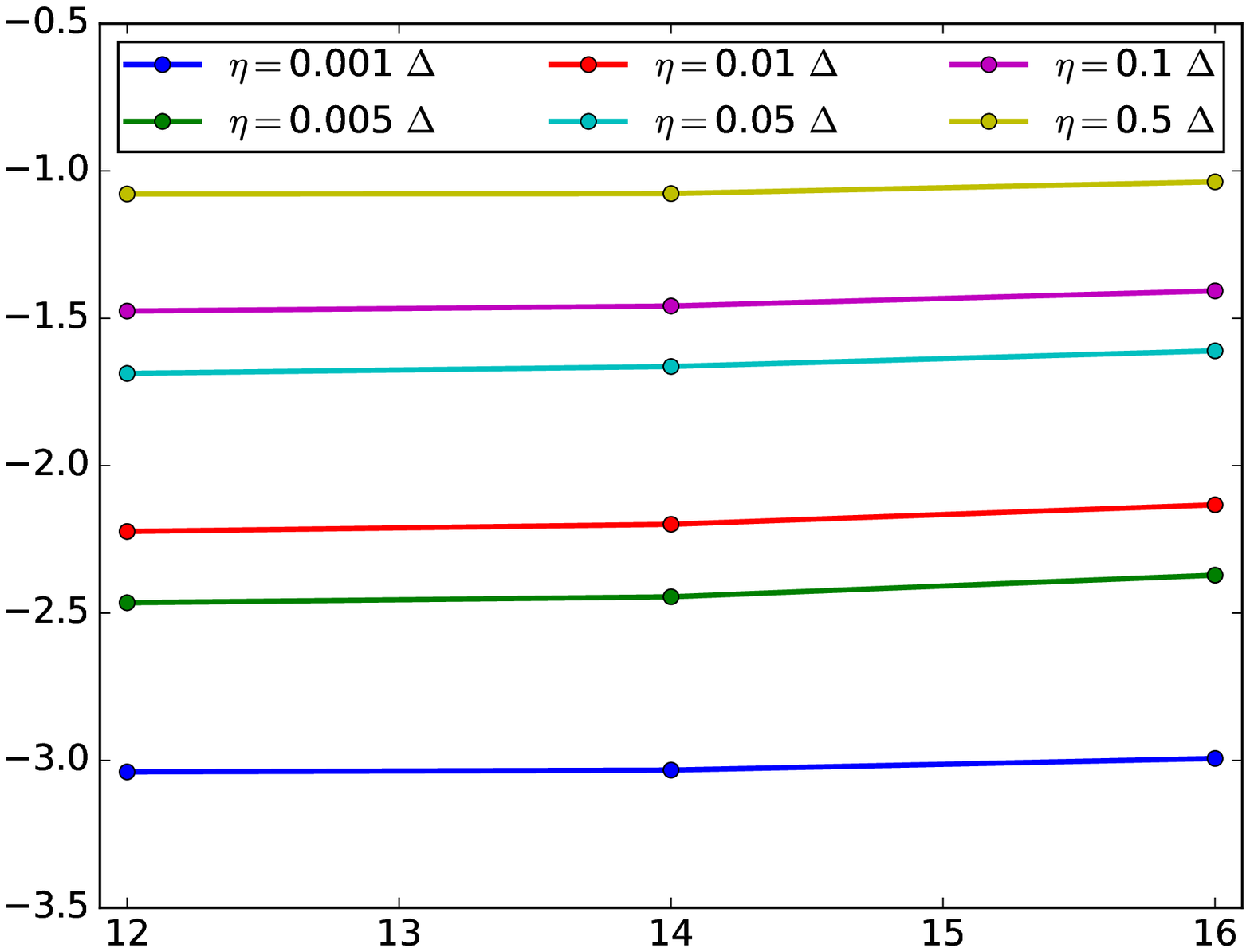}};
	\node (img2) [below= of img1,node distance=0cm,xshift=0cm, yshift=1.60cm]{\includegraphics[width=0.5\textwidth]{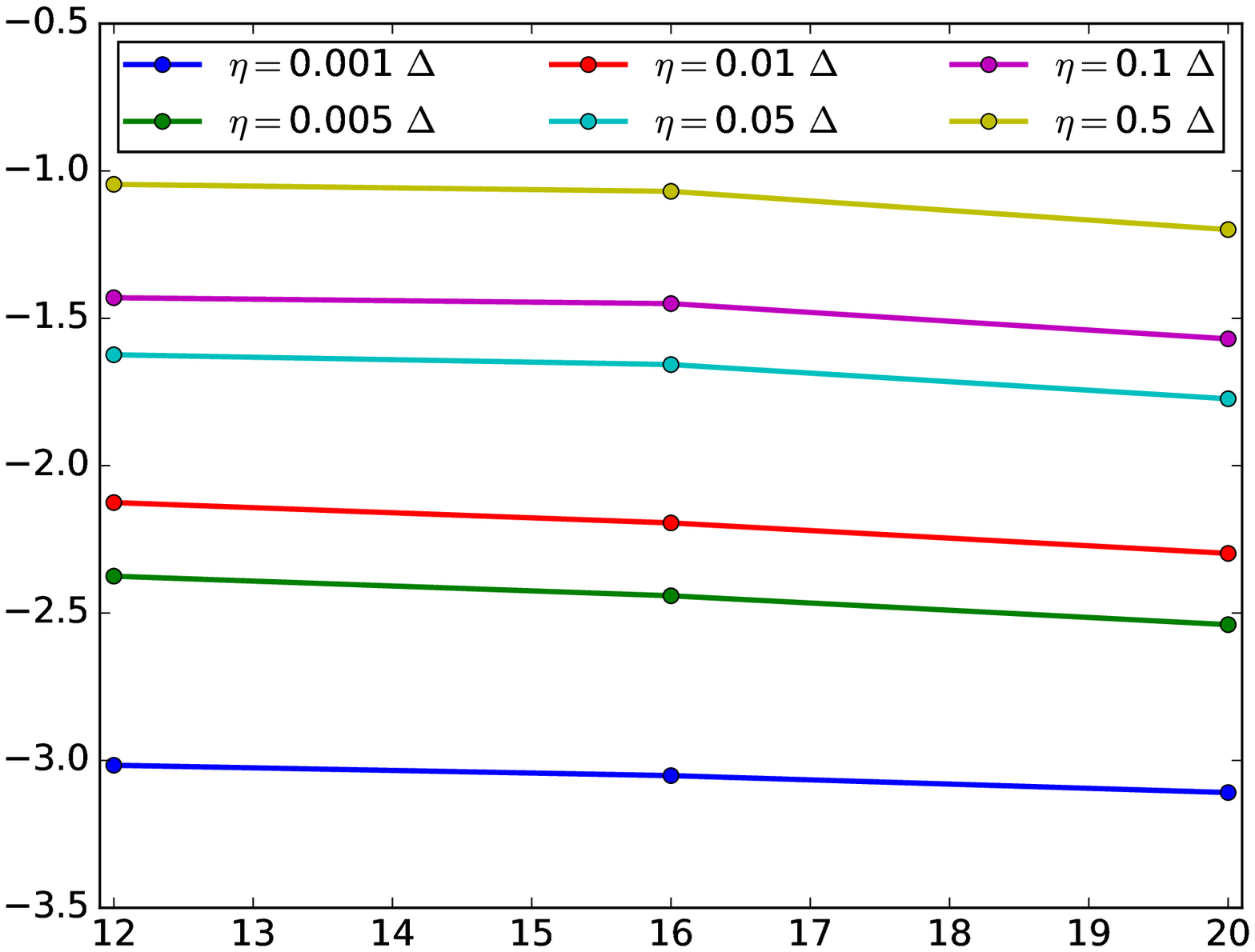}};
	\node[left=of img1,node distance=0cm,yshift=1.0cm,xshift=1.4cm,rotate=90]{$log(\sigma_{dc})$};
		\node[left=of img2,node distance=0cm,yshift=1.0cm,xshift=1.4cm,rotate=90]{$log(\sigma_{dc})$};
%	\node[below=of img1,node distance=0cm,yshift=1.2cm]{System size $L$};
	\node[below=of img2,node distance=0cm,yshift=1.4cm]{System size $L$};

	\node[above= of img1,node distance=0cm,yshift=-7.25cm,xshift=-2.65cm]{\fbox{$\nu=1/2$}};
	\node[above= of img2,node distance=0cm,yshift=-7.25cm,xshift=-2.65cm]{\fbox{$\nu=1/4$}};
	\end{tikzpicture}
	\caption{Variation of dc conductivity with system size for different choices of Lorentzian width: This shows the variation of $\sigma_{dc}$ vs. system size curve for different choices of width $\eta$ of the Lorentzian at filling fractions $\nu=1/2$(Top) and $\nu=1/4$(Bottom). Again $\eta$ is chosen as fraction $f$ of the mean level spacing $\Delta$ where $f$ is changed from $0.001$ to $1.0$ for both the cases.}
	\label{dc-eta}
\end{figure}

Here we calculate the dc conductivity for different filling fractions $\nu=\frac{1}{2},\frac{1}{4}$ at $h=0.6, V=1.0, t=1.0$ and look for the system size scaling of the same.  The sample sizes chosen are $108000,108000,18000,1800$ for $L=10, 12, 14, 16$ respectively at $\nu=1/2$ and $108000,108000,900$ for $L=12,16,20$ respectively at $\nu=1/4$. Since for $\nu=1/6$ we have only two system sizes accessible($L=18,24$) through exact diagonalization, we do not investigate the scaling of the dc conductivity in this case.

Fig.~\ref{dc} shows the variation of $log(\sigma_{dc})$ with increasing system sizes at different filling fractions $\nu$. While for $\nu=1/2$(fig~\ref{dc} top panel) the $\sigma_{dc}$ remains roughly of same order with increasing system size, for $\nu=1/4$(fig~\ref{dc} bottom panel) we see decay with increasing system size which is slower than that in the MBL phase where the decay is expected to be exponential~\cite{dc_cond_setiawan}. This behavior is further investigated in section-IV where we calculate the contribution of three different energy regions (thermal, non-ergodic extended and MBL) to the $\sigma_{dc}$ separately.
\paragraph*{Choice of width $\eta$:}
Since we are working in the regime $\omega \rightarrow 0$, changing the width $\eta$ of the Lorentzian corresponding to $\delta(E_n-E_m)$ implies changing the number of energy pairs $(E_n,E_m)$ contributing to the $\sigma_{dc}$. Therefore decreasing $\eta$ leads to decrease in the number of energy pairs contributing to $\sigma_{dc}$ and hence decrease in $\sigma_{dc}$. However, since $\eta$ is chosen as a fraction of the corresponding mean level spacing $\Delta$, the nature of the variation of $\sigma_{dc}$ with $L$ does not change. Only the absolute value of the dc conductivity changes with $\eta$  as seen in fig.~\ref{dc-eta}.
The sample sizes chosen for this calculation are $(102400,6400,128)$ for $L=12,14,16$ respectively in case of $\nu=1/2$, $(10240,10240,128)$ for $L=12,16,20$ respectively for filling fraction $\nu=1/4$.
\section{Level spacing statistics, ETH violation, Entanglement entropy:}
To identify the non-ergodic extended phase we employ different diagnostics which differentiate between non-ergodic and ergodic phases, and extended and localized phases.
\paragraph{Level Spacing statistics:}
The level spacing statistics shows the presence or absence of repulsion in the energy spectrum. Level repulsion is generally associated with the ergodicity of the system and its absence with the non-ergodicity. It was argued that for a localized system, two eigenstates having similar energies are far apart in terms of overlap and hence do not experience any level repulsion. As a result, successive energy gaps become Poisson distributed~\cite{OganesyanHuse.2007}.
On the other hand, in the ergodic phase, there is level repulsion and the level spacing statistics is that of a random matrix theory (RMT) specifically the Gaussian orthogonal ensemble(GOE) for the systems we consider. To identify the two different types of behavior, we calculate the average of the ratio of successive energy gaps $r_n=\frac{min\{\delta_n,\delta_{n+1}\}}{max\{\delta_n,\delta_{n+1}\}}$ where $\delta_n=E_{n+1}-E_n$ is the energy gap between the $n^{th}$ and $(n+1)^{th}$ energy level. It is known that the average value of this ratio is $\simeq0.529$ for GOE while for a Poisson distribution it is $\simeq0.386$.
Thus, the average level spacing ratio can be used to locate the ergodic and non-ergodic phases.
Here, we divide the whole many body energy spectrum into equal bins and calculate the average level spacing ratio for the successive eigenstates belonging to the same energy bin.
While calculating the level spacing ratio, we average over $10000,5000,200$ realizations of the random phase $\phi$ for the system sizes $L=12,14,16$ respectively for $\nu=1/2$, $10000,4800,96$ realizations of the random phase $\phi$ for the system sizes $L=12,16,20$ respectively for $\nu=1/4$ and $9600,100$ realizations of the random phase $\phi$ for system sizes $L=18,24$ respectively for $\nu=1/6$.

\begin{figure*}[t]
  \centering
  \begin{tikzpicture}
    \node (img1){\includegraphics[width=0.35\textwidth]{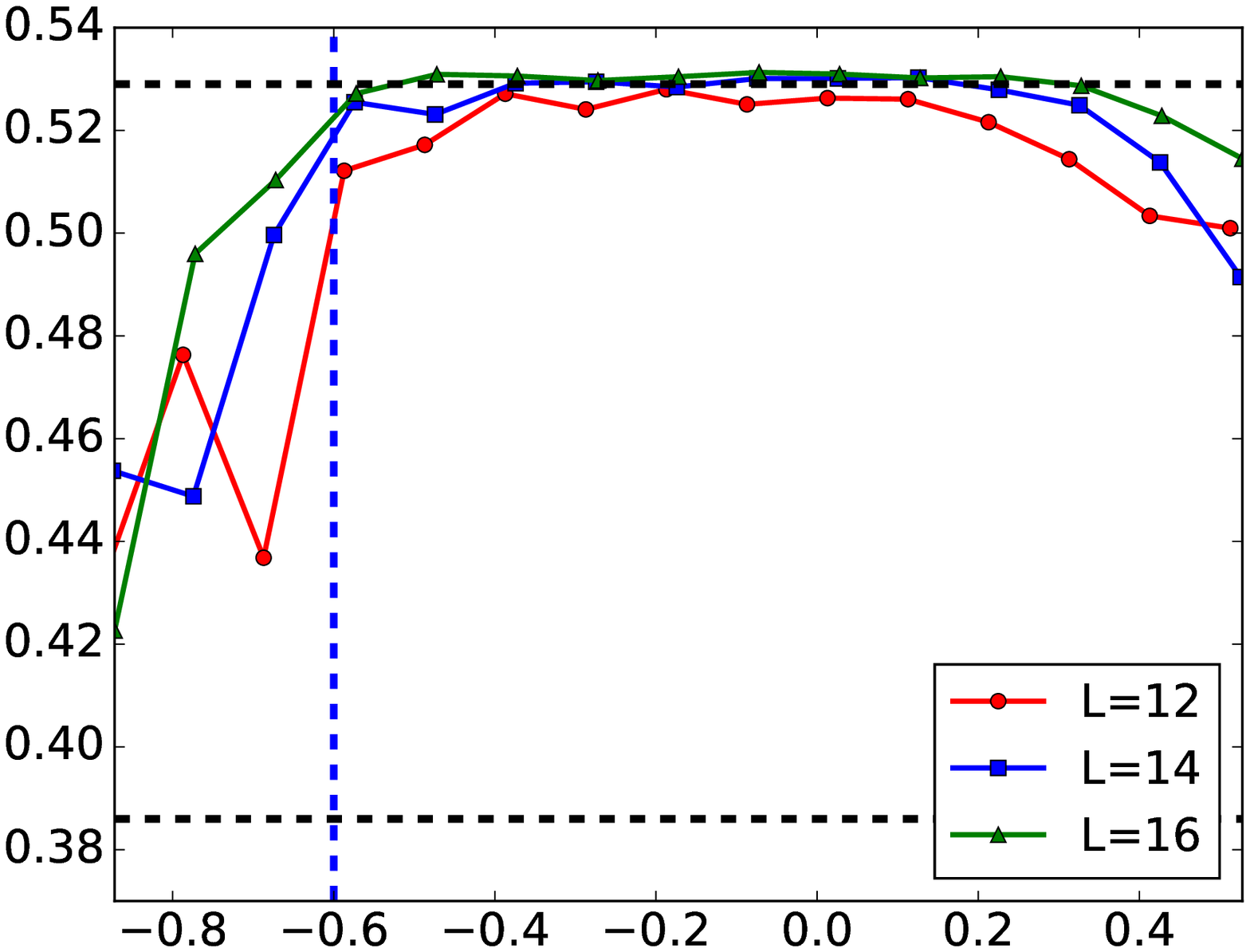}};
    \node (img2) [right= of img1,node distance=0cm,xshift=-1.5cm]{\includegraphics[width=0.35\textwidth]{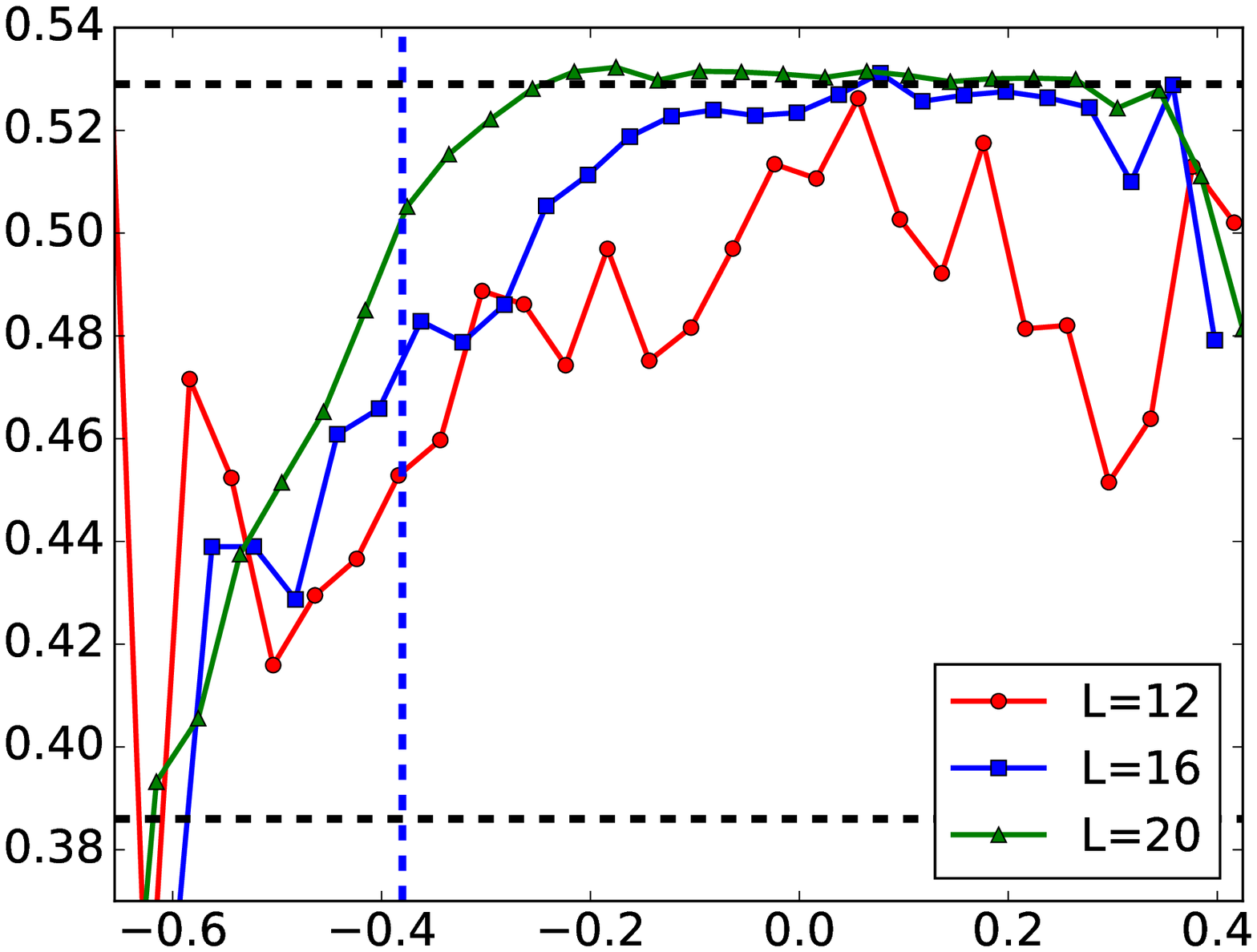}};
    \node (img3) [right= of img2,node distance=0cm,xshift=-1.5cm]{\includegraphics[width=0.35\textwidth]{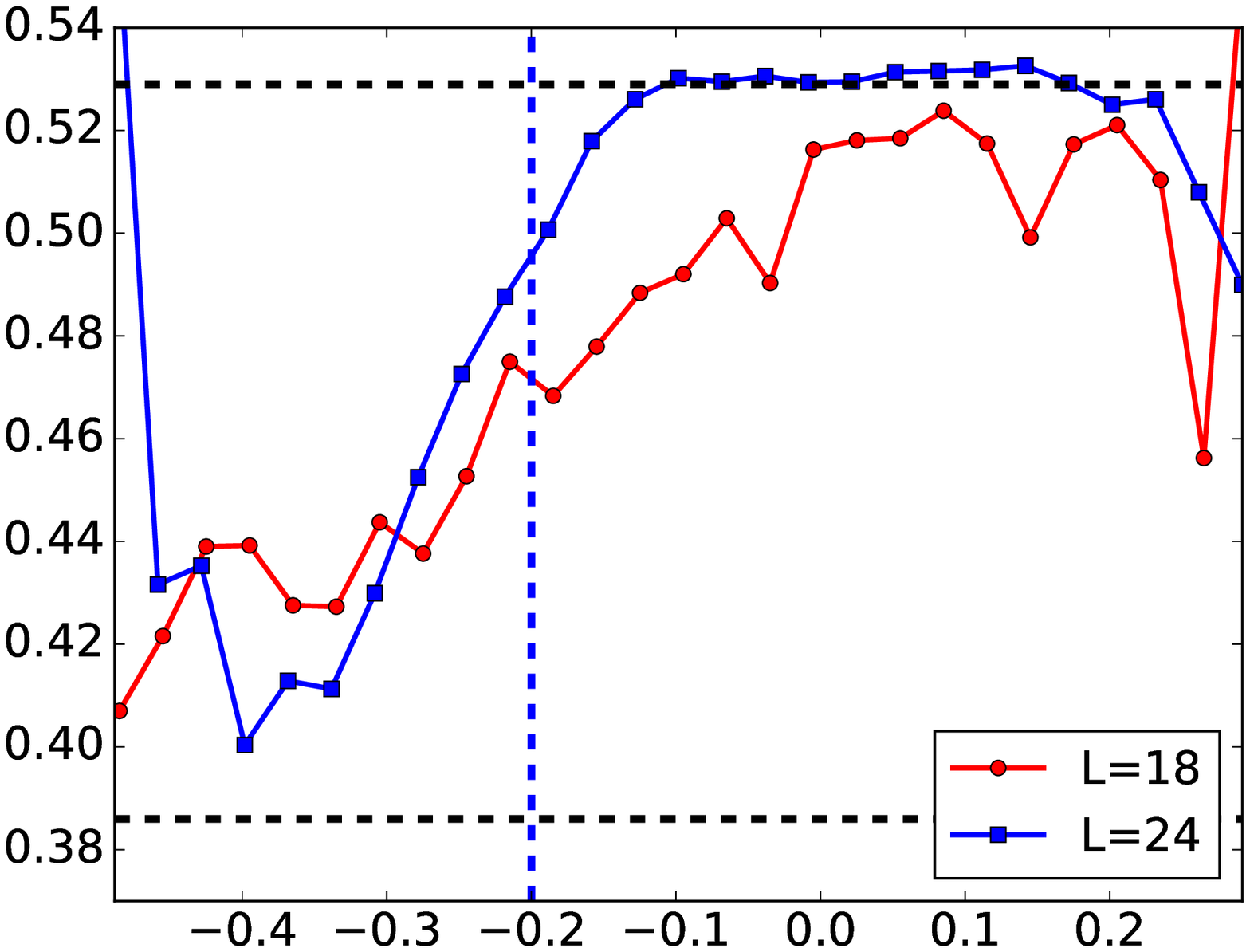}};
    \node (img4) [below= of img1,node distance=0cm,xshift=0.0cm,yshift=1.5cm]{\includegraphics[width=0.35\textwidth]{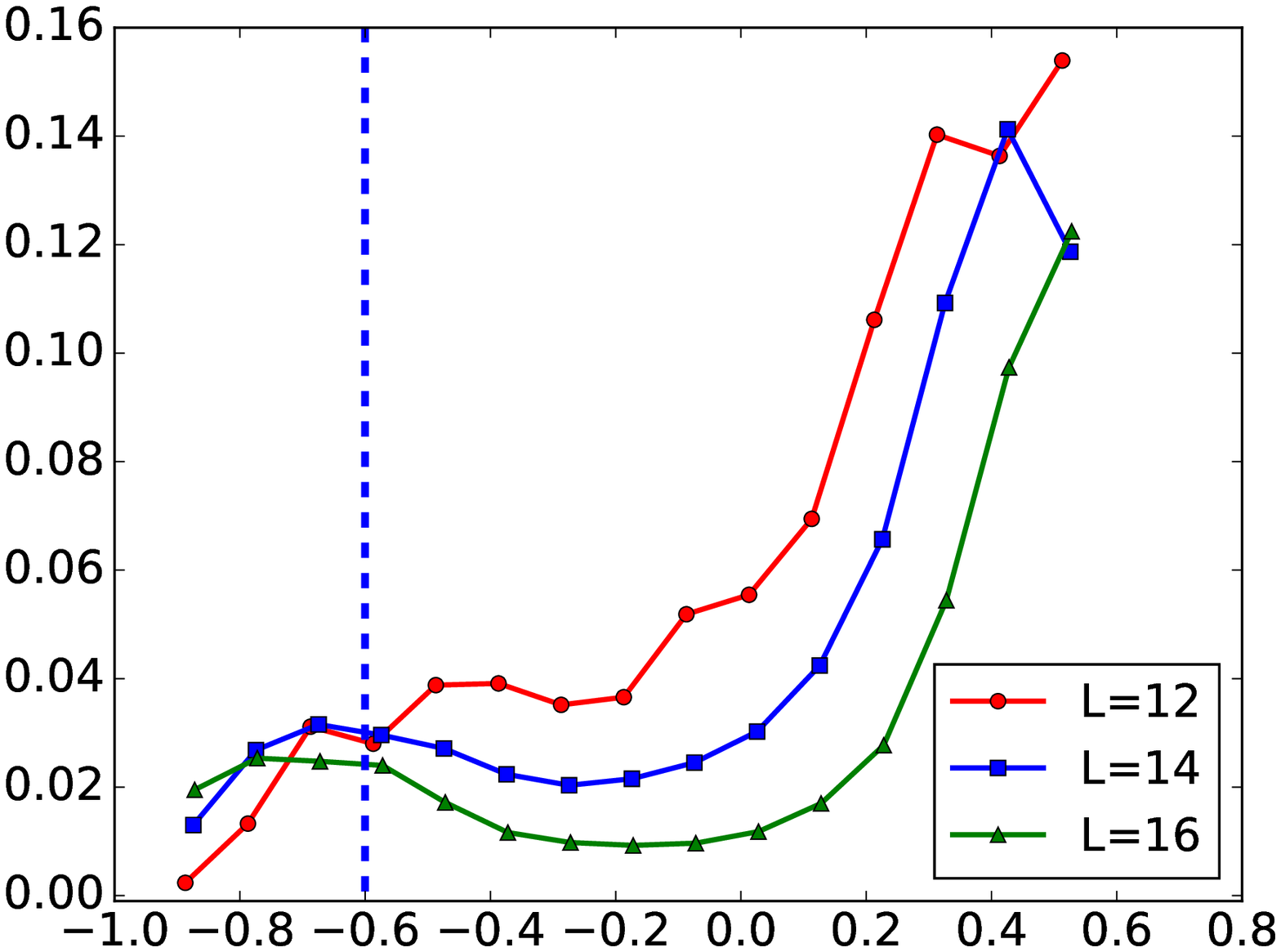}};
    \node (img5) [below= of img2,node distance=0cm,xshift=0.0cm,yshift=1.5cm]{\includegraphics[width=0.35\textwidth]{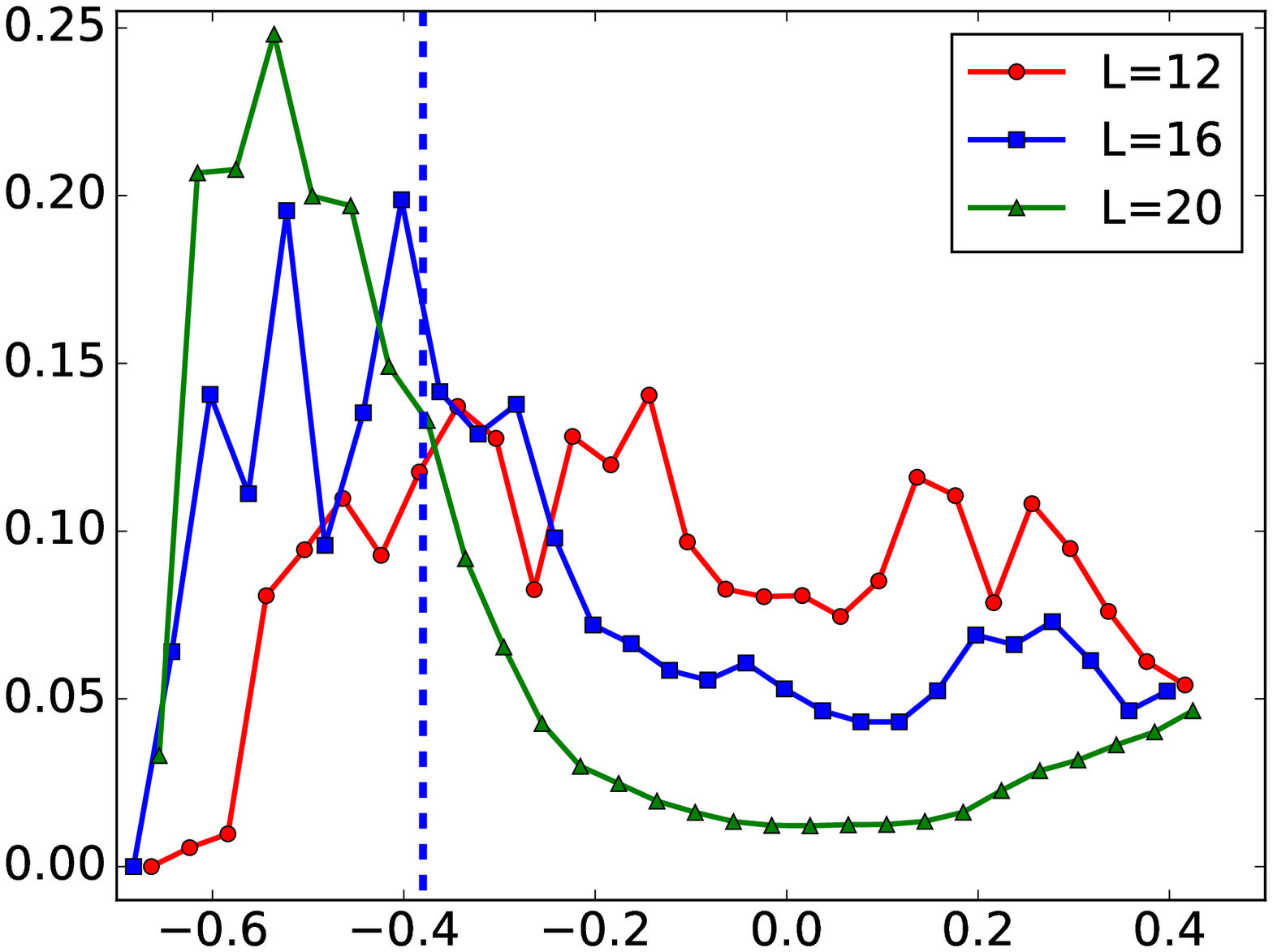}};
    \node (img6) [below= of img3,node distance=0cm,xshift=0.0cm,yshift=1.5cm]{\includegraphics[width=0.35\textwidth]{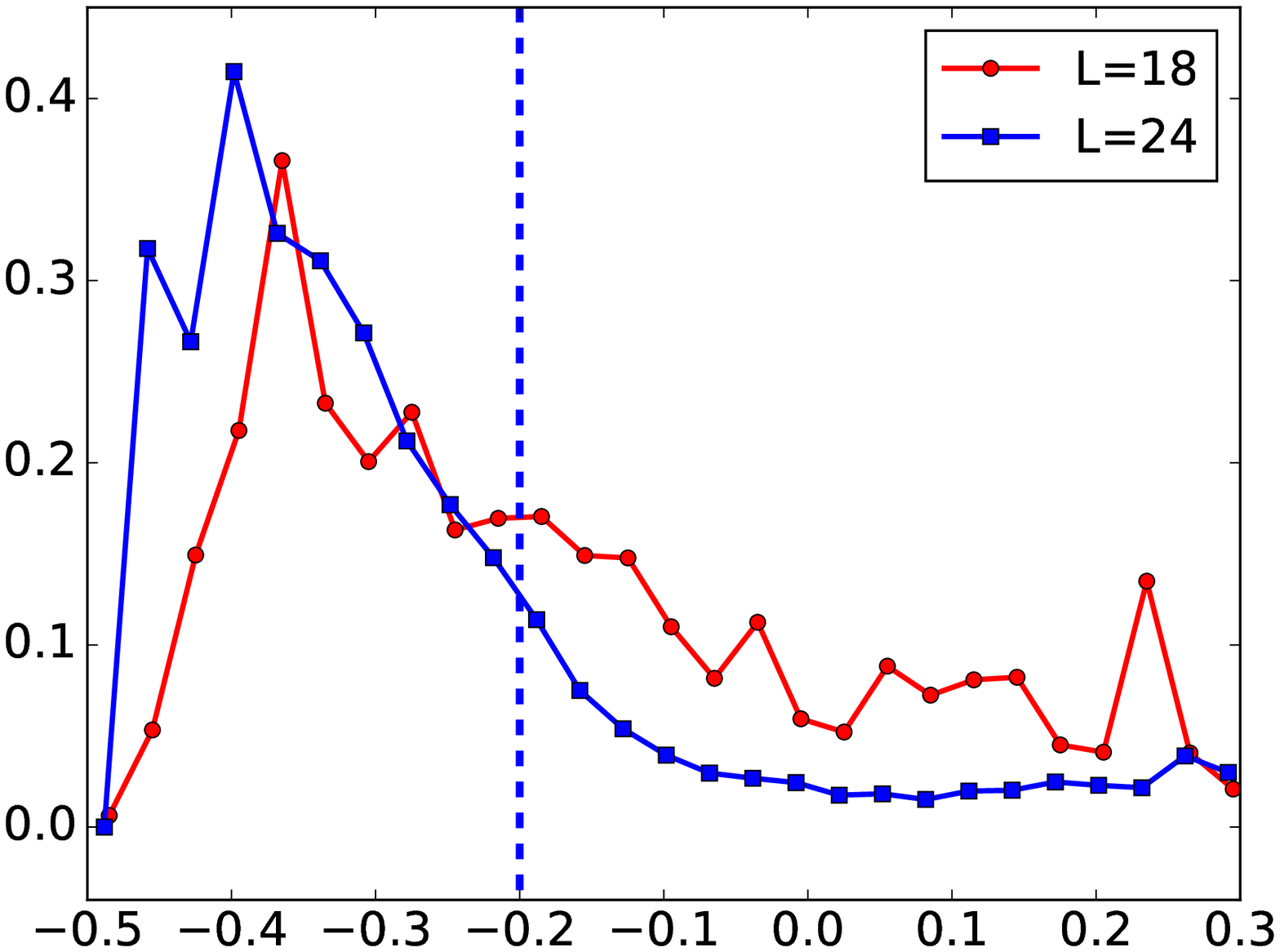}};
   \node (img7)[below=of img4,node distance=0cm,xshift=0.0cm,yshift=1.5cm]{\includegraphics[width=0.35\textwidth]{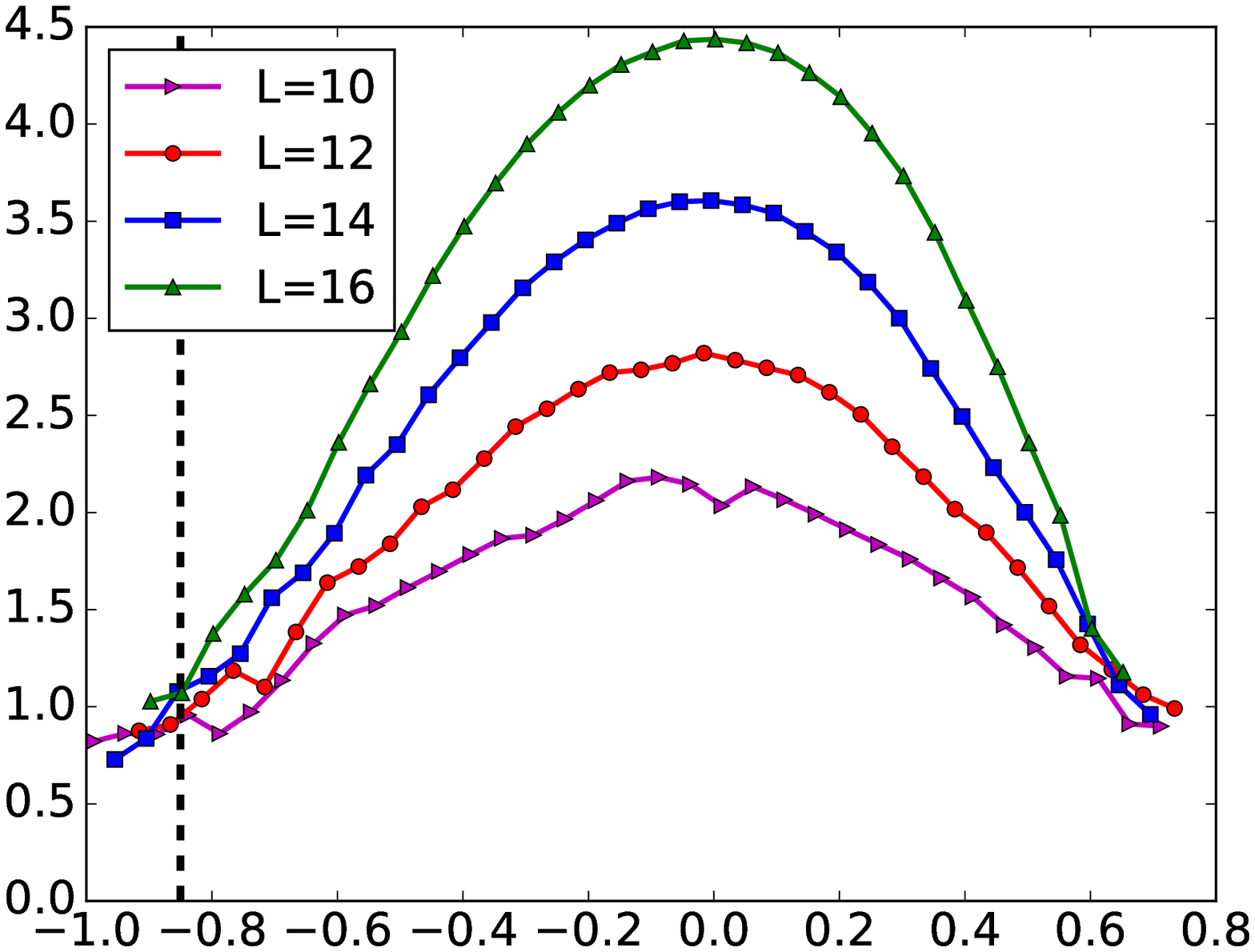}};
    \node (img8) [below= of img5,node distance=0cm,xshift=0.0cm,yshift=1.5cm]{\includegraphics[width=0.35\textwidth]{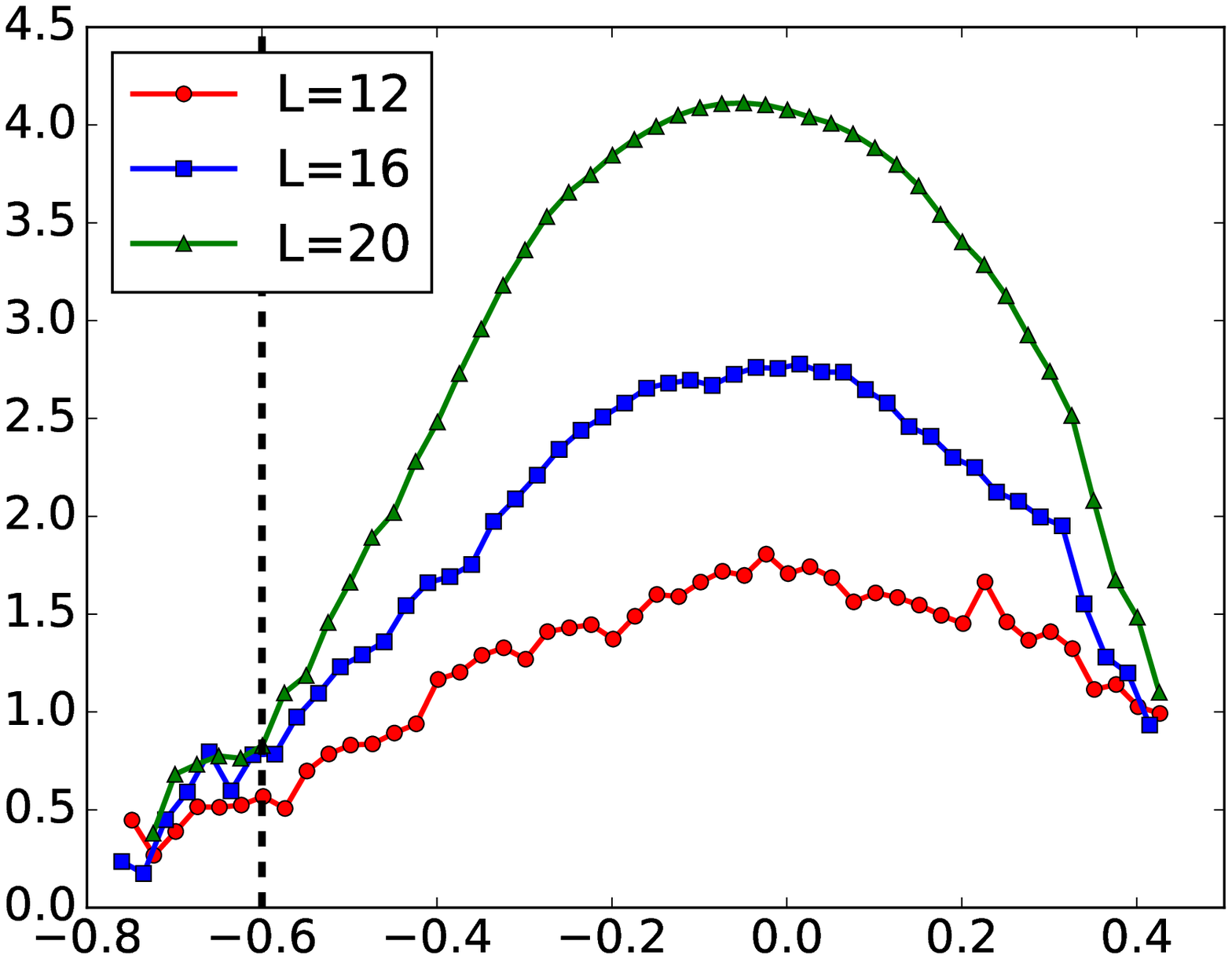}};
    \node (img9) [below= of img6,node distance=0cm,xshift=0.0cm,yshift=1.5cm]{\includegraphics[width=0.35\textwidth]{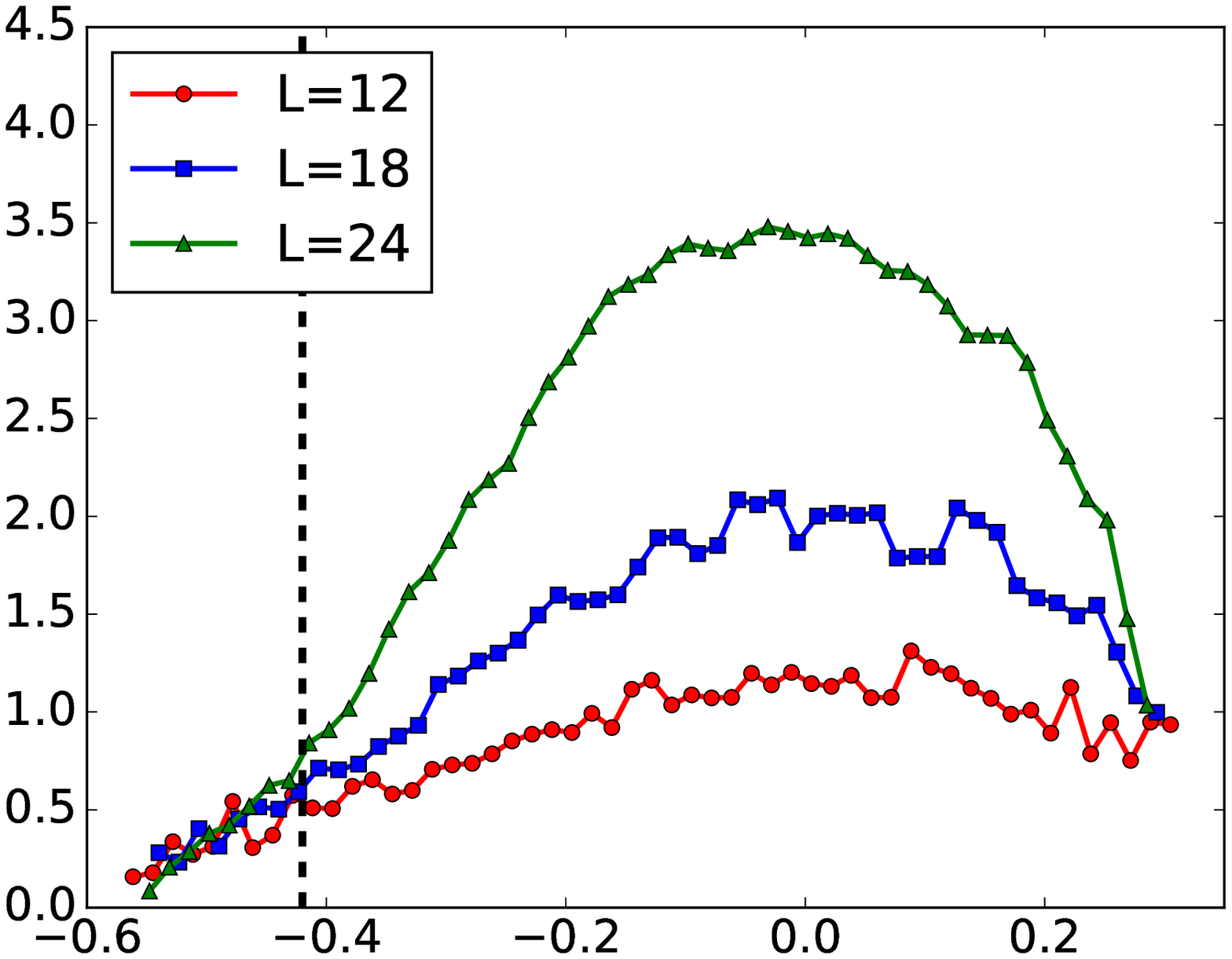}};
   \node[left=of img1,node distance=0cm,yshift=0.5cm,xshift=1.0cm,rotate=90]{$<r>$};
    \node[left=of img4,node distance=0cm,yshift=1.2cm,xshift=1.0cm,rotate=90]{Variance $var[O]$};
        \node[left=of img7,node distance=0cm,yshift=1.5cm,xshift=1.0cm,rotate=90]{R\'enyi entropy $S_2$};   
   \node[below=of img7,node distance=0cm,yshift=1.2cm]{Energy density $\varepsilon$};
   \node[below=of img8,node distance=0cm,yshift=1.2cm]{Energy density $\varepsilon$};
   \node[below=of img9,node distance=0cm,yshift=1.2cm]{Energy density $\varepsilon$};
   
\node[above= of img1,node distance=0cm,yshift=-2.3cm,xshift=-2.0cm]{$(a)$};
	\node[above= of img2,node distance=0cm,yshift=-2.3cm,xshift=-2.0cm]{$(b)$};
	\node[above= of img3,node distance=0cm,yshift=-2.3cm,xshift=-2.0cm]{$(c)$};   
	
	\node[above= of img4,node distance=0cm,yshift=-2.3cm,xshift=-2.0cm]{$(d)$};
	\node[above= of img5,node distance=0cm,yshift=-2.3cm,xshift=-2.0cm]{$(e)$};
	\node[above= of img6,node distance=0cm,yshift=-2.3cm,xshift=-2.0cm]{$(f)$};   
	
	\node[above= of img7,node distance=0cm,yshift=-2.3cm,xshift=2.0cm]{$(g)$};
	\node[above= of img8,node distance=0cm,yshift=-2.3cm,xshift=2.0cm]{$(h)$};
	\node[above= of img9,node distance=0cm,yshift=-2.3cm,xshift=2.0cm]{$(i)$};   
   
    \node[above= of img1,node distance=0cm,yshift=-3.8cm,xshift=1.0cm]{\fbox{$\nu=1/2$}};
        \node[above= of img2,node distance=0cm,yshift=-3.8cm,xshift=1.0cm]{\fbox{$\nu=1/4$}};
        \node[above= of img3,node distance=0cm,yshift=-3.8cm,xshift=1.0cm]{\fbox{$\nu=1/6$}};

%\node[below=of img1,node distance=0cm,yshift=1.2cm]{Energy density $\varepsilon$};
  \end{tikzpicture}
  \caption{Level spacing statistics, ETH violation and entanglement entropy: The average value of $r_{n}$ at different energies obtained by dividing the spectrum into bins centered around those energies and location of $E_{r}$ indicated by blue dashed line (as mentioned in section III a.) for different filling fractions is shown: $\nu=\frac{1}{2}~(a),\frac{1}{4}~(b),\frac{1}{6}~(c)$. The horizontal black dashed lines in $(a)$, $(b)$ and $(c)$ depict the average value of the ratio for GOE($\simeq 0.529$) and Poisson($\simeq 0.386$) distributions. The variance of the observable $O$ obtained by dividing the spectrum into bins and location of $E_{O}$ indicated by the blue vertical dashed line (as mentioned in section III b.) for different filling fractions is shown: $\nu=\frac{1}{2}~(d),\frac{1}{4}~(e),\frac{1}{6}~(f)$. The bipartite R\'enyi entropy and location of $E_{S}$ indicated by the black vertical dashed line (as mentioned in section III c.) for different filling fractions is shown: $\nu=\frac{1}{2}~(g),\frac{1}{4}~(h),\frac{1}{6}~(i)$.}
  \label{fig-lsr-eth}
\end{figure*}

As seen in fig.~\ref{fig-lsr-eth}[a,b,c] there is a value of energy $E_r$ (blue dashed line) for all filling fractions up to which all the energy eigenstates display Poisson level spacing statistics implying the absence of level repulsion in the systems. Above this value of the energy density all the energy eigenstates display GOE statistics implying ergodicity in the system.

\paragraph{ETH violation:}
The eigenstate thermalization hypothesis (ETH)~\citep{deutsch.1991,srednicki.1994,Rigol.2008} states that in an ergodic system the expectation value of a local operator in any many body eigenstate $\ket{n}$ corresponds to the thermal expectation value at the corresponding energy $E_n$. In other words, any single energy eigenstate gives the same expectation value of a local operator as the one calculated by averaging over nearby energy eigenstates, i.e. the thermal expectation value corresponding to microcanonical ensemble. The fluctuation of the expectation value in nearby energy  eigenstates is exponentially small in the system size. 
Therefore if we define an observable such as the total number of particles in one half of the system,
\begin{equation}
\nonumber
O(E_n)=\sum \limits_{j=1}^{N/2} \braket{n|n_j|n}
\end{equation}
then $O(E_n)$ should be similar for nearby many body energy eigenstates if the system is ergodic\citep{LiGaneshan.2015} and a fluctuation of the values of $O(E_n)$ should imply non-ergodicity.

However, for finite size systems there are always finite fluctuations in $O(E_n)$ between adjacent energy eigenstates for both ergodic and non-ergodic phases. Hence, we distinguish between the two phases by looking at the fluctuation in the quantity $O(E_n)$ among nearby energy eigenstates, in an ergodic phase where ETH holds, this is exponentially small in the system size whereas in a non-ergodic phase, it is of order one.

We divide the many-body spectrum into bins and calculate the variance of $O(E_n)$ within each bin. The quantity is averaged over the same number of samples for the offset $\phi$ as in the level spacing ratio calculation. Fig.~\ref{fig-lsr-eth}[d,e,f] shows the variance of $O(E_n)$ within nearby energies as a function of energy density. The value of $var[O]$ is independent of system size up to an energy density equal to $E_O$ indicating a non-ergodic phase for all filling fractions up to that energy. For larger energy densities, $var[O]$ gets suppressed with increasing system size indicating an ergodic phase. We also note that $E_r \simeq E_O$.
\paragraph{Entanglement entropy:}
To study the localization-delocalization transition, here we study the bipartite R\'enyi entropy between the two halves of the system.\citep{renyi_entropy, LiGaneshan.2015} For that we divide the lattice of size $L$ in two halves, $A$ and $B$ of size $L/2$ and compute the R\'enyi entropy $S_{2}$ using
\begin{equation}
\nonumber
S_{2}(L/2)=-\log(\Tr\rho_{A}^{2})
\end{equation}
where $\rho_{A}$ is the reduced density matrix obtained from the density matrix $\rho=\ket{\Psi_E} \bra{\Psi_E}$ by taking a partial trace over subsystem $B$. Here $\Psi_E$ is the many body eigenstate corresponding to many body energy $E$.

A volume law behavior of the entanglement entropy implies delocalized states while area law behavior implies localization in the system. Here we calculate the entanglement entropy as a function of energy density by binning the many-body energy spectrum and calculating the average R\'enyi entropy for each bin. While calculating the entropy, we average over $100,100,100, 10$ realizations of the random phase $\phi$ for the system sizes $L=10,12,14,16$ respectively for $\nu=1/2$, and $50,20,10$ realizations of the random phase $\phi$ for the system sizes $L=12,16,20$ respectively for $\nu=1/4$ and $1000,100,15$ realizations of the random phase $\phi$ for system sizes $L=12,18,24$ respectively for $\nu=1/6$.

Fig~\ref{fig-lsr-eth}[g,h,i] shows the variation of R\'enyi entropy $S_2$ as a function of energy density. There is an energy density value $E_S$ (marked by the dashed line) which separates the many body energy eigenstates following area law behavior on the left of $E_S$ from the volume law following eigenstates on the right.

We see that the energy density $E_r$ that separates the energy levels following a Poissonian level spacing distribution from the ones following GOE is almost equal to the energy density $E_O$ separating the energy eigenstates violating ETH from those that obey ETH. We also see the energy density $E_S$ that separates the localized eigenstates from delocalized eigenstates is less than $E_r \simeq E_O$. Therefore, we have three different regions for all three filling fractions considered: localized and non-ergodic (energy density less than $E_S$), delocalized and non-ergodic (energy density in between $E_S$ and $E_r$) and delocalized and ergodic (energy density larger than $E_r\simeq E_O$). Note that an ergodic phase necessarily has delocalized states  since they are  essential for diffusive transport which is a defining characteristic of ergodicity.

\section{Energy resolved dc conductivity}
The presence of $\delta(E_n-E_m)$ in eqn.~\ref{eq-dc} implies that states with essentially equal energies $(E_n\simeq E_m)$ (due to the broadening of the delta function) contribute $|\braket{m|J|n}|^2$ to $\sigma_{dc}$. Therefore, one can study the contribution of the three phases, many-body localized (region-I), non-ergodic and extended (region-II) and ergodic (region-III) separately to the dc conductivity.
\begin{figure}[t]
	\centering
	\begin{tikzpicture}
	\node (img1) [right= of img1,node distance=0cm]{\includegraphics[width=0.5\textwidth]{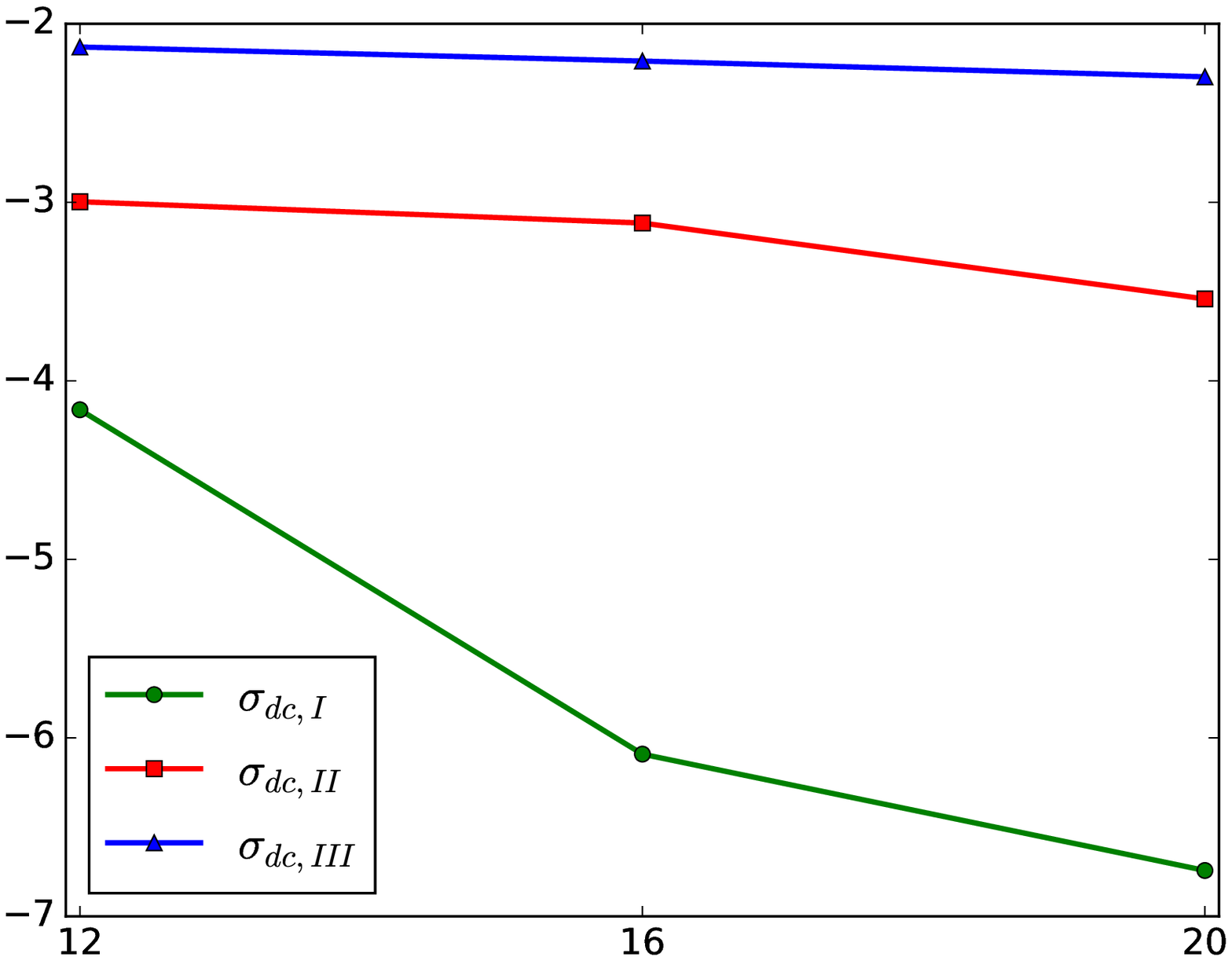}};
	\node[left=of img1,node distance=0cm,yshift=1.0cm,xshift=1.5cm,rotate=90]{$log(\sigma_{dc,n})$};
	\node[below=of img1,node distance=0cm,yshift=1.6cm]{System size $L$};	
	\end{tikzpicture}
	\caption{Energy-resolved dc conductivity: Variation of contributions of the three phases to $\sigma_{dc}$ ($\sigma_{dc,I}$, $\sigma_{dc,II}$ and $\sigma_{dc,III}$) with increasing system sizes are shown for filling fraction $\nu= \frac{1}{4}$. The width $\eta$ is chosen to be $0.01\Delta$ where $\Delta$ is the mean level spacing.}
	\label{fig-dc-part}
\end{figure}
\begin{equation}
\nonumber
\sigma_{dc}\simeq \sigma_{dc,I}+\sigma_{dc,II}+\sigma_{dc,III}
\end{equation}
Fig.~\ref{fig-dc-part} shows the variation of $log(\sigma_{dc,n})$ as a function of the system size where $n=I,II,III$ denote the contribution from the three different regions for $\nu=1/4$. Here again we average $\log(\sigma_{dc,n})$ over 96000, 18000 and 900 randomly chosen offset angle $\phi$'s for the system sizes 12, 16 and 20 respectively. As seen in fig.~\ref{fig-dc-part} the $\sigma_{dc,I}$ appears to decay much faster with system size as compared to $\sigma_{dc,II}$ and $\sigma_{dc,III}$. In contrast $\sigma_{dc,II}$ undergoes a slower decay with the system size while $\sigma_{dc,III}$ hardly decays at all with increasing system sizes. The diagnostics of the previous section show that the region-I has properties of MBL phase,  namely Poisson distributed energy levels, area law scaling of the entanglement entropy and ETH violation. It is known that in the MBL phase the dc conductivity decreases exponentially with the increasing system size and goes to zero in the thermodynamic limit.~\citep{dc_cond_mbl_1, dc_cond_mbl_2, dc_cond_setiawan}. Therefore the slower decay of $\sigma_{dc,II}$, which corresponds to the non-ergodic delocalized phase, compared to that in the region-I (the MBL phase) suggests that the non-ergodic extended phase has a dc conductivity which in the thermodynamic limit goes to zero, but more slowly than in the MBL phase. This is perhaps indicative of sub-diffusive transport (i.e. power law decay of conductivity with the system size) in the non-ergodic delocalized phase.

\section{Discussion and Conclusions}
 We have calculated the ac and dc conductivities (fig.~\ref{ac}, \ref{ac-eta}, \ref{dc}, \ref{dc-eta}) and energy-resolved level spacing statistics (fig.~\ref{fig-lsr-eth}[a,b,c]) to understand the effects of the non-ergodic extended phase on transport. We identified the non-ergodic (ETH violating like the many body localized eigenstates) yet extended (with volume-law scaling of entanglement entropy like the thermal eigenstates) eigenstates for different filling fractions (fig.~\ref{fig-lsr-eth}).

 Looking at the energy-resolved level spacing, we find that the average value of the level spacing ratio $r_n$ changes as a function of energy density from the Poisson value ($\big \langle r_{n}\big \rangle_{Poisson}\simeq 0.386$) to the GOE value ($\big \langle r_{n}\big \rangle_{GOE}\simeq 0.5295$) and that there is energy density $E_r$ which separates these two regions. We also notice that the transition from Poissonian to GOE statistics is concurrent with the transition between non-ergodic to ergodic phase ($E_O$) as identified by analyzing validity of ETH as a function of energy density. The energy density $E_s$ corresponding to transition between area law entangled eigenstates and volume law entangled eigenstates is less than $E_O$ or $E_r$ for all cases. Therefore the energy level statistics captures the non-ergodic to ergodic transition as opposed to the localized to delocalized transition as captured by entanglement entropy. This is consistent with what happens in other systems for example inthe  context of Anderson model in Bethe lattice~\citep{LucaScardicchio.2014,bethelattice.2012} where for weak disorder, the system is in extended and non-ergodic phase and have Poisson level statistics.

In our studies of conductivity (fig.~\ref{ac}, \ref{dc}), we have observed signatures of sub-diffusive transport- (i) the exponent, $\alpha$, obtained from the log-log plots of $\sigma(\omega)$ vs $\omega$ lies  between 0 and 1 and (ii)  vanishing dc conductivity where the dc conductivity appears to go to zero in the thermodynamic limit slower than in the MBL phase where the conductivity decreases exponentially with increasing system sizes. Studying the contributions of the three different kinds of eigenstates to the dc conductivity (fig. \ref{fig-dc-part}), we observe that the contribution of the ergodic states is indicative of diffusive transport where it hardly decays with increasing system sizes. On the other hand, the conductivity in the non-ergodic localized phase appears to decay much faster with increasing system sizes while in the non-ergodic extended phase it appears to decay at a rate in between the two cases. As the non-ergodic localized phase (MBL phase) is expected to have an exponential decay of the DC conductivity with increasing system size, a slower decay of the DC conductivity contribution from the non-ergodic extended phase is perhaps suggestive of a power law decay of the same with system size. A similar power law decay was also observed~\citep{dc_cond_setiawan} for systems with random disorder where the sub-diffusive phase arises due to the Griffiths effect. We thus provide evidence suggestive of sub-diffusive transport even in the presence of a deterministic quasiperiodic potential which does not allow for Griffiths-regions (which occur in the presence of random disorder), which we attribute to the presence of the `non-ergodic extended' phase.

\appendix*
\onecolumngrid
\section{System size dependence of $\sigma(\omega)$}
\begin{figure*}[h]
	\centering
	\begin{tikzpicture}
	\node (img1){\includegraphics[width=0.35\textwidth]{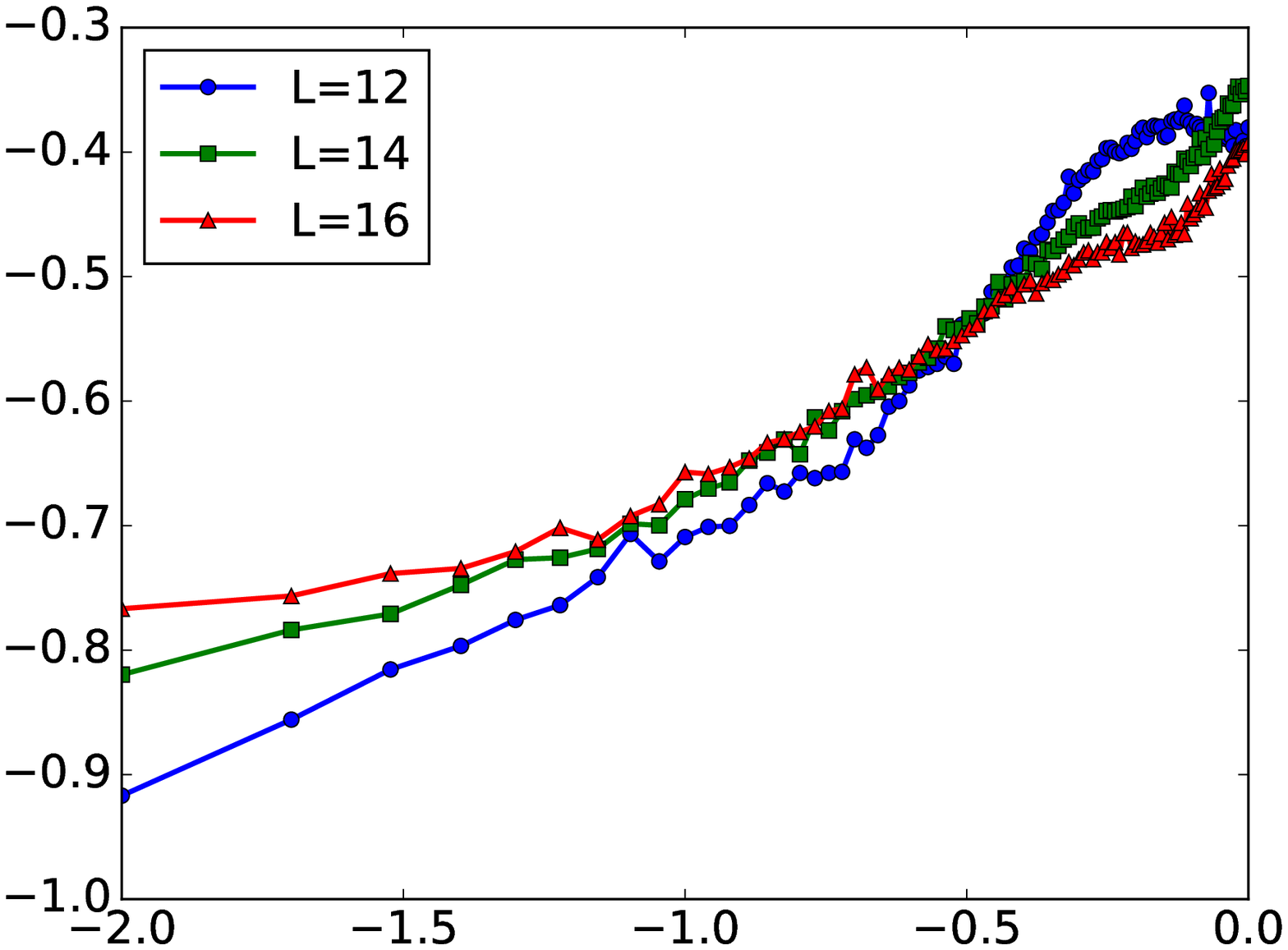}};
	\node (img2) [right= of img1,node distance=0cm,xshift=-1.5cm] {\includegraphics[width=0.35\textwidth]{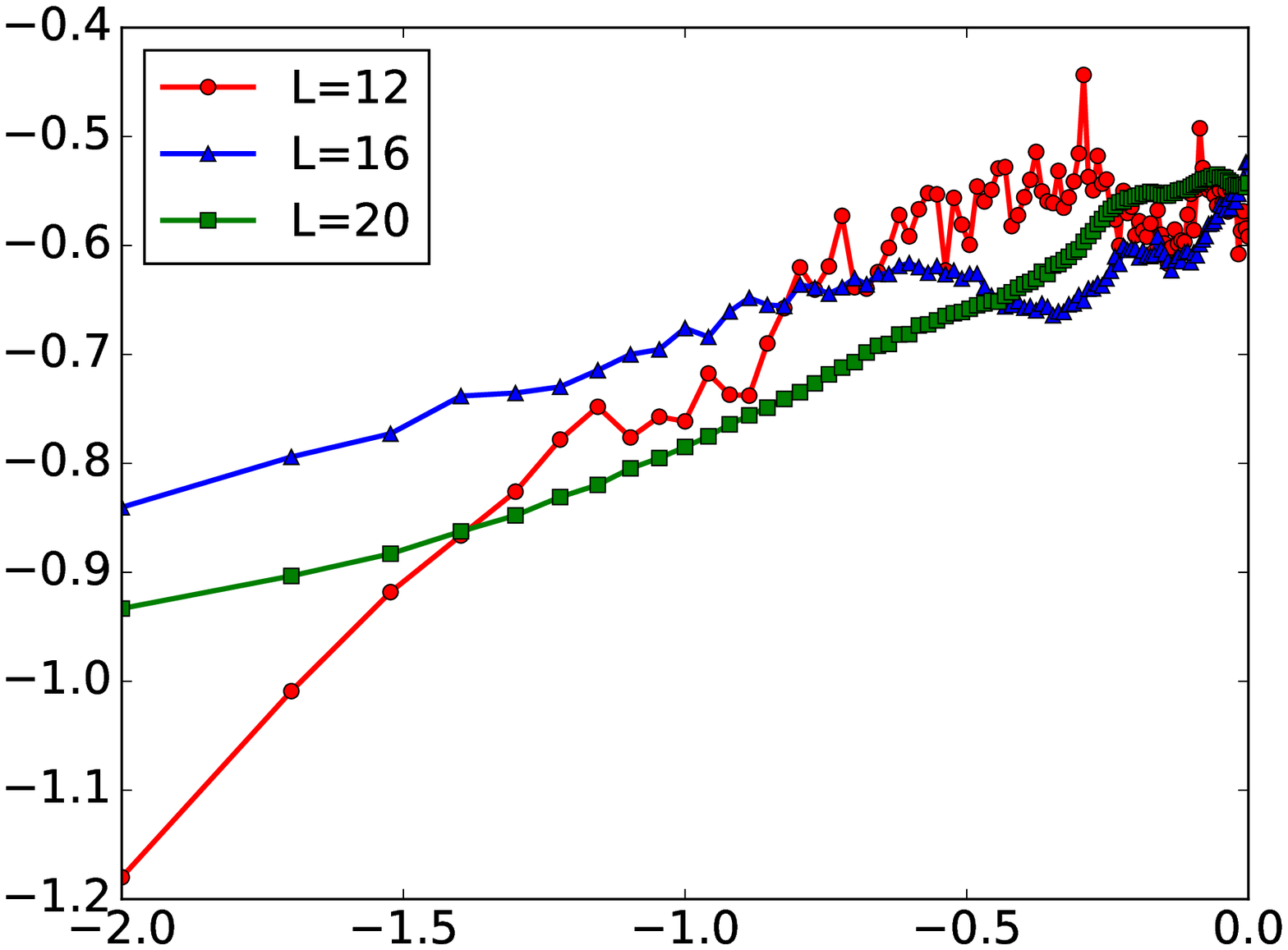}};
	\node(img3) [right= of img2, node distance=0cm,xshift=-1.5cm] {\includegraphics[width=0.35\textwidth]{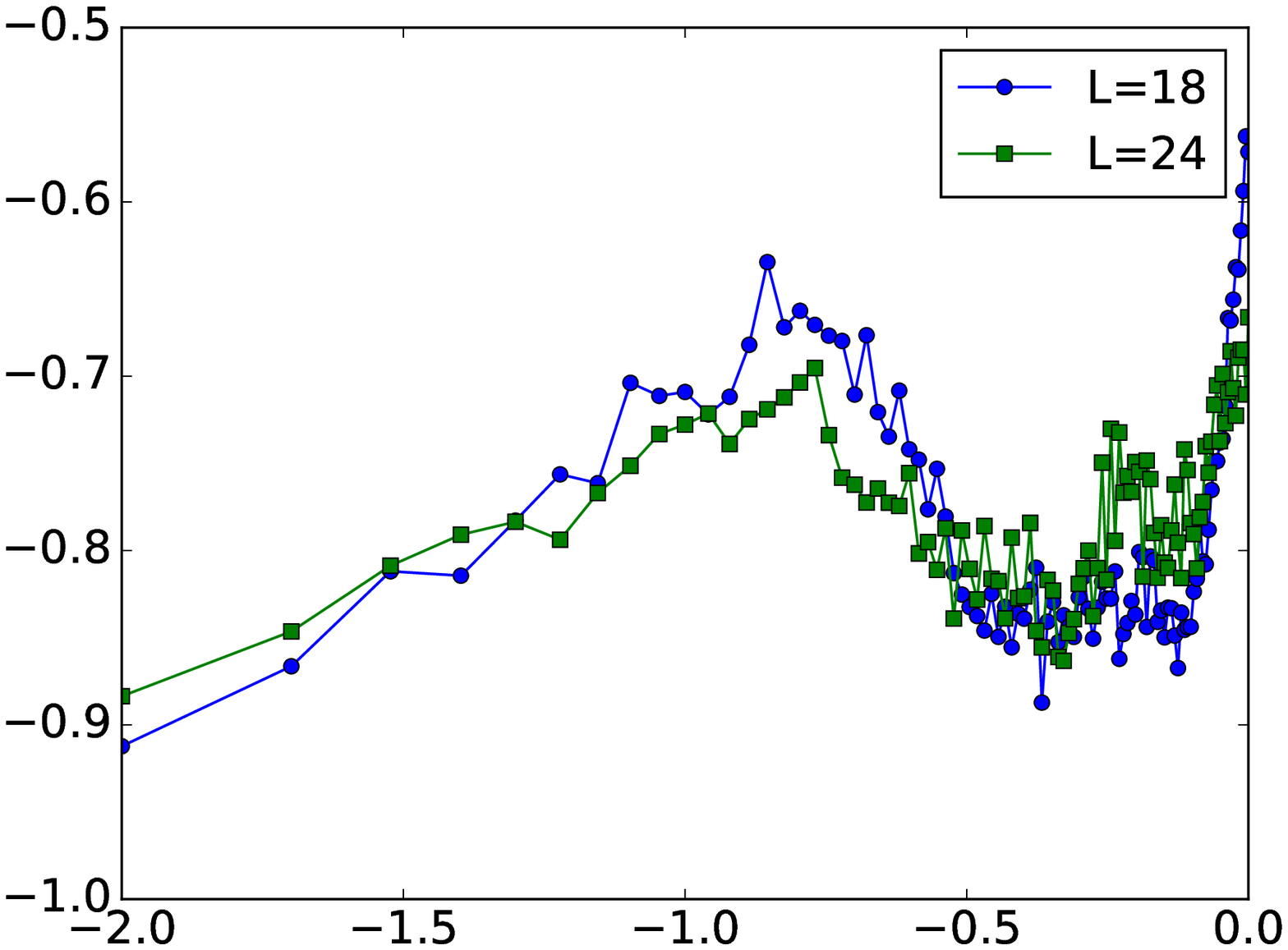}};
	\node[left=of img1,node distance=0cm,yshift=1.0cm,xshift=1cm,rotate=90]{$log\left[\sigma(\omega)\right]$};
	\node[below=of img1,node distance=0cm,yshift=1.2cm]{$log(\omega)$};
	\node[below=of img2,node distance=0cm,yshift=1.2cm]{$log(\omega)$};
	\node[below=of img3,node distance=0cm,yshift=1.2cm]{$log(\omega)$};

	\node[above= of img1,node distance=0cm,yshift=-2.3cm,xshift=0.0cm]{$\nu=1/2$};
	\node[above= of img2,node distance=0cm,yshift=-2.3cm,xshift=0.0cm]{$\nu=1/4$};
	\node[above= of img3,node distance=0cm,yshift=-2.3cm,xshift=0.0cm]{$\nu=1/6$};
	\end{tikzpicture}
	\caption{$\sigma(\omega)$ vs $\omega$ plot for different system sizes: This plot shows the variation $\sigma(\omega)$ vs. $\omega$ curve for different system sizes at different filling fractions. The chosen sample sizes are $(4800,640,128)$ for $L=12,14,16$ respectively at $\nu=\frac{1}{2}$, $(6400,4800)$ for $L=12,16$ respectively at $\nu=\frac{1}{4}$ and  $(4800,64)$ for $L=18,24$ respectively at $\nu=\frac{1}{6}$. The width $\eta$ of the Lorentzian is chosen $0.001\Delta$ for all cases where $\Delta$ is the mean level spacing.}
	\label{ac-system}
\end{figure*}
\twocolumngrid
Fig.~\ref{ac-system} shows the system size dependence of $\sigma(\omega)$ versus $\omega$ curves at different filling fractions. Here we see the dependency of $\sigma(\omega)$ on $\omega$ is almost independent of system size for fillings $\nu=1/2$ and $1/6$, for $\nu=1/4$ there is a large dependency on the system size. While for $L=12,\:\nu=1/4$ the ac conductivity for the considered range of $\omega$ depends strongly on the choice of broadening $\eta$, the other two system sizes $L=16,20$ do not have any $\eta$ dependency. Therefore, it is not clear why there is this system size dependency in the $\sigma(\omega)$ vs. $\omega$ curves for $L=16$ and $20$. We also see that the same effect also occurs for Aubry Andre model ($h_i=h cos(2\pi \alpha i+\phi)$) at filling $\nu=1/4$. This needs further investigation.

%\bibliography{reference}

%merlin.mbs apsrev4-1.bst 2010-07-25 4.21a (PWD, AO, DPC) hacked
%Control: key (0)
%Control: author (72) initials jnrlst
%Control: editor formatted (1) identically to author
%Control: production of article title (-1) disabled
%Control: page (0) single
%Control: year (1) truncated
%Control: production of eprint (0) enabled
%

\end{document}